\documentclass[a4paper,11pt]{article}
\title{Energy-based Modelling of the Feedback Control of Biomolecular Systems with Cyclic Flow Modulation}
\pdfoutput=1 
\usepackage[noblocks]{authblk}
\author[1,2]{Peter J. Gawthrop\footnote{Corresponding
    author. \textbf{peter.gawthrop@unimelb.edu.au}}}
\affil[1]{
  Systems Biology Laboratory,
  Department of Biomedical Engineering,
  Melbourne School of Engineering,
  University of Melbourne,
  Victoria 3010, Australia.
   }
   \affil[2]{Systems Biology Laboratory,
     School of Mathematics and Statistics,
     University of
      Melbourne University of Melbourne, Victoria 3010}

\usepackage{abstract}
\usepackage{a4wide,times}
\usepackage[numbers,compress]{natbib}
\usepackage{siunitx}
\newcommand{\Si}[1]{(\si{#1})}

\usepackage[hidelinks]{hyperref}
\usepackage{url,doi}

\usepackage[section]{placeins}
\usepackage{graphicx,subfigure,fancybox,color}

\newcommand{\BG}[1]{\text{\sffamily\textbf{#1}}}

\newcommand{\Ce}{\BG{Ce }}

\newcommand{\one}{\BG{1 }}
\newcommand{\zero}{\BG{0 }}
\newcommand{\One}{\BG{1}}
\newcommand{\Zero}{\BG{0}}

\renewcommand{\Re}{\BG{Re }}

\newcommand{\BGL}[2]{$\BG{#1}$:$\mathbf{#2}$} 

\newcommand{\BCe}[1]{\BGL{Ce}{#1}}

\newcommand{\BR}[1]{\BGL{R}{#1}}

\newcommand{\BRe}[1]{\BGL{Re}{#1}}

\usepackage{amsmath,amssymb,amscd,amstext,mathtools,extarrows,centernot,bm}

\newcommand{\jw}{j\omega}

\newcommand{\where}{\text{where }}
\newcommand{\hence}{\text{hence }}

\newcommand{\ddt}[1]{\frac{d}{dt}{#1}}

\newcommand{\NN}{N}

\newcommand{\xx}{x}

\renewcommand{\AA}{{A}}

\newcommand{\ff}{f}

\newcommand{\Css}{C}
\newcommand{\Dss}{D}

\newcommand{\xxc}{\xx_{ch}}
\newcommand{\xtc}{\xt_{ch}}
\newcommand{\xbc}{\xb_{ch}}

\usepackage{listings}
\newcommand{\xt}{\tilde{x}}
\newcommand{\xb}{\bar{x}}
\newcommand{\xtb}{\bar{\xt}}

\newcommand{\ft}{\tilde{f}}
\newcommand{\fb}{\bar{f}}

\newcommand{\tfs}[3]{{#1}_{#2}(#3)}
\newcommand{\tf}[2]{#1(#2)}
\newcommand{\Ls}{\tf{L}{s}}
\newcommand{\Las}{\tfs{L}{act}{s}}
\newcommand{\Lps}{\tfs{L}{pas}{s}}

\newcommand{\Lw}{\tf{L}{\jw}}
\newcommand{\Gs}[1]{\tfs{G}{#1}{s}}

\usepackage{chemformula}

\newcommand{\Ch}[1]{(\ch{#1})}

\usepackage{graphicx,subfigure,fancybox,color}
\graphicspath{{./Figs/}}

\newcommand{\Fig}[2]{
 \includegraphics[width=#2\linewidth]{#1.pdf}
}

\newcommand{\SubFig}[3]{
 \subfigure[#2]{
   \includegraphics[width=#3\linewidth]{#1.pdf}
   \label{subfig:#1}
 }
}

\begin{document}
\maketitle
\begin{abstract}
  Energy-based modelling brings engineering insight to the
  understanding of biomolecular systems. It is shown how
  well-established control engineering concepts, such as loop-gain,
  arise from energy feedback loops and are therefore amenable to
  control engineering insight. In particular, a novel method is
  introduced to allow the transfer function based approach of
  classical linear control to be utilised in the analysis of feedback
  systems modelled by network thermodynamics and thus amalgamate energy-based
  modelling with control systems analysis.
  The approach is illustrated using a class of metabolic cycles with
  activation and inhibition leading the concept of Cyclic Flow
  Modulation.
\end{abstract}
\tableofcontents
\newpage
\section{Introduction}
\label{sec:introduction}
The bond graph implementation of Network Thermodynamics was introduced
some 50 years ago as an energy-based approach to modelling
biomolecular systems \citep{OstPerKat71,OstPerKat73}. ``Graphical representations
similar to engineering circuit diagrams can be constructed for
thermodynamic systems. ... such diagrams do increase one's intuition
about system behaviour.''\citep{Per75}. 

The design of linear feedback circuits also has a long history
and the correspondingly well-established theory of control systems
\citep{AstMur08} has been applied to biomolecular systems
\citep{Sav09,CloWel10,VecNinSon08} and has led to a number of control
concepts such as feedback and integral action being used in the
biomolecular context \citep{YiHuaSi00,ChaArmPap13,MuzGomMet09,QiaVec18}.
%

Classical linear control theory is based on \emph{transfer function}
models of dynamical systems.  In contrast, the energy-based approach
of this paper uses the bond graph paradigm for modelling biomolecular
systems. There has been limited work on the bond graph approach to
control \cite{Kar79a,ShaHogHar91,Gaw95c}
For this reason, a novel method is introduced to allow the transfer
function based approach of classical linear control to be utilised in
the analysis of feedback systems modelled by bond graphs and thus
combine energy-based modelling with control systems analysis.
This approach can, in principle, be used in two distinct ways: to
\emph{analyse} existing  biomolecular control systems arising from evolution
and to \emph{synthesise} new biomolecular control systems for
synthetic biology.

As discussed by \citet{GawCra16}, the bond graph approach gives the set
of \emph{nonlinear} ordinary differential equations describing the
biomolecular system being modelled.  Linearisation of non-linear
systems is a standard technique in control engineering.
%
Linearisation in the context of bond graph models of biomolecular
systems was introduced by \citet{GawCra16} and is used here.


The role of metabolic cycles in the regulation of metabolic flux is
well established \citep{NewChaCra84,QiaBea06,Cor12,ChaArmPap13,GarGri17}. Such cycles%
\footnote{ The pejorative term ``futile cycle'' is often used to
  describe such cycles; this will be avoided in this paper.  }
are involved in a number of substrate conversions including those
between
fructose-6-phosphate and fructose-1,6-biphosphate,
fructose-6-phosphate and fructose-2,6-biphosphate,
triglyceride/fatty
acid, glucose and glucose-6-phosphate,
and glycogen and glucose 1-phosphate \citep{NewChaCra84,Cor12,GarGri17}.
To illustrate the fusion of network thermodynamics and control theory,
this paper will focus on the first two inter-conversions involving
fructose-6-phosphate \Ch{F6P}.
Because of the cyclic nature of these two reactions, and the fact that
flow is modulated, the term \emph{Cyclic Flow Modulation} (CFM) is
used to describe such reaction systems.

The use of CFM requires energy and there is a trade-off between
quality of control and energy consumed~\cite{QiaBea06}. It is
therefore important to account for energy flows when modelling
biomolecular systems and this is done here using the fusion of the
network thermodynamics paradigm, as implemented using bond graphs,
with control theory.
%
Criteria for robust biochemical reaction networks have been
established which ensure zero steady-state error
\citep{ShiFei11,AokLilGup19,KimEnc20}; but these papers make no
mention of energy and therefore entirely ignore thermodynamic
constraints.

Building complex systems is simplified using modularity~\cite{NeaCooSmi14}; but it is
essential to distinguish two different concepts of modularity:
\emph{computational} modularity where physical correctness is retained and
\emph{behavioural} modularity where module behaviour (such as
ultra-sensitivity) is retained~\cite{GawCra16}. As well as providing computational
modularity, bond graphs provide a natural formulation of behavioural
modularity and reveal the sources of retroactivity~\cite{GawCra16}.
\emph{Chemostats} \cite{GawCra16,PolEsp14} are used to create an open
system from a closed system and also provide a convenient way of
providing \emph{ports} to connect bond graph  modules.

\S~\ref{sec:bond-graph-based-1} provides a brief tutorial introduction
to the bond graph  approach in this context.
\S~\ref{sec:bond-graph-based} introduces the bond graph based
approach to the analysis of feedback control systems using an
enzyme catalysed reaction with competitive inhibition as an
illustrative example.
\S~\ref{sec:CFM} shows how cyclic flow modulation (CFM) can be used to
build effective feedback controllers with approximate integral action.
\S~\ref{sec:conclusion} concludes the paper and gives directions for
future work.

\subsection{Bond graph based modelling of reaction systems}
\label{sec:bond-graph-based-1}
\begin{figure}[htbp]
  \centering
  \SubFig{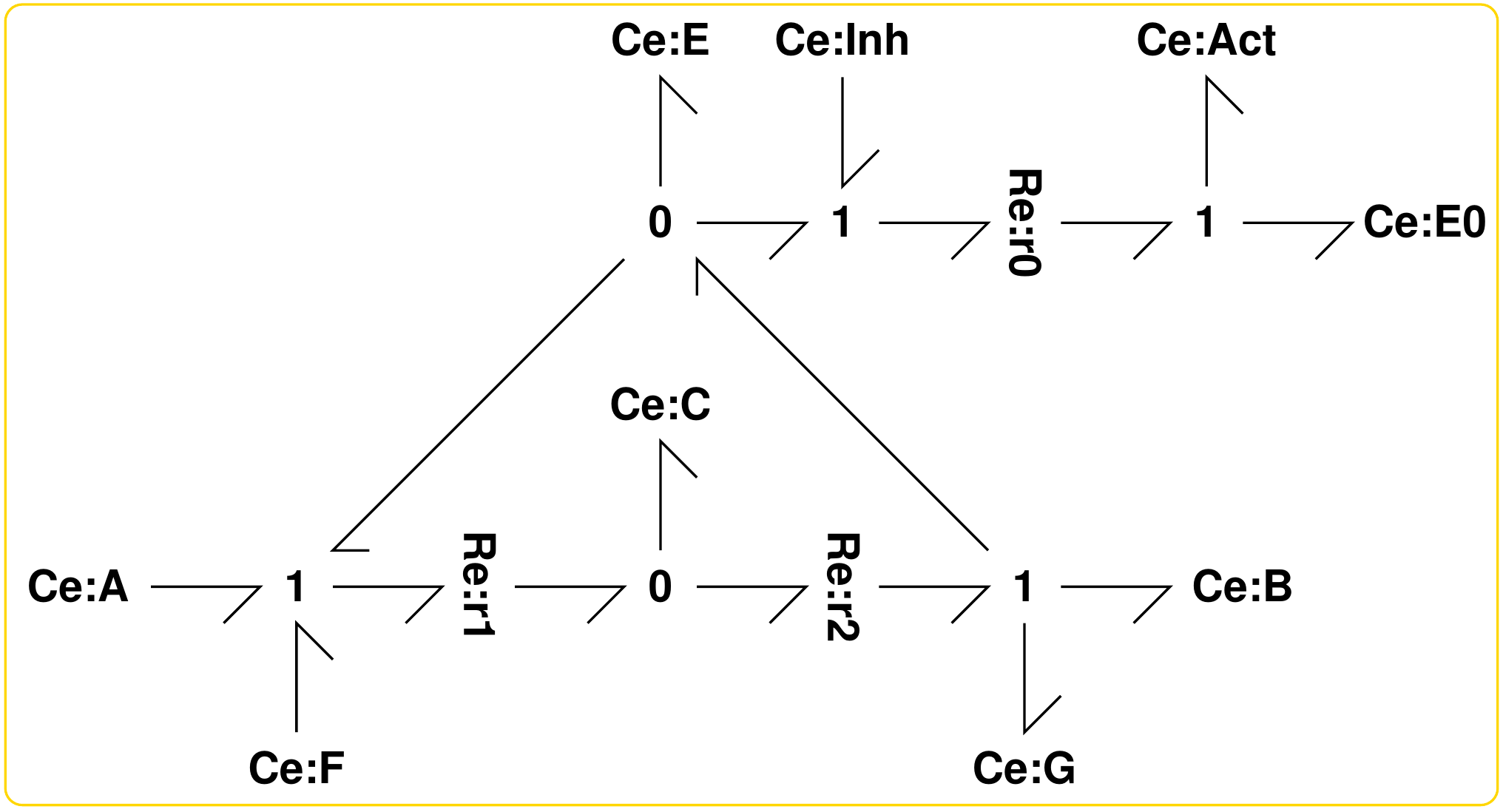}{Enzyme-catalysed reaction (module \textbf{ecr})}{0.6}\\
  \SubFig{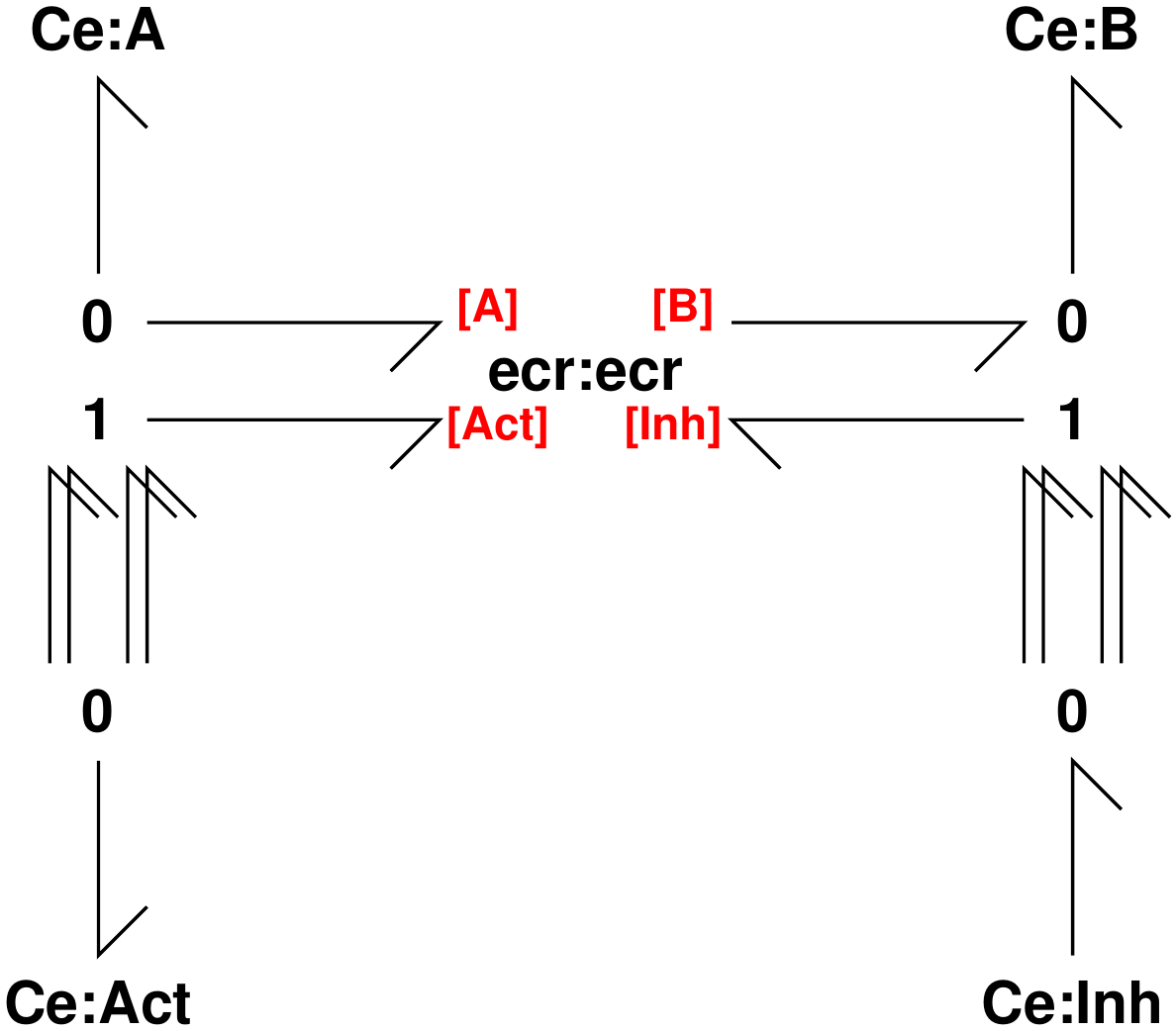}{Cooperativity (module \textbf{ECR})}{0.4}
  \SubFig{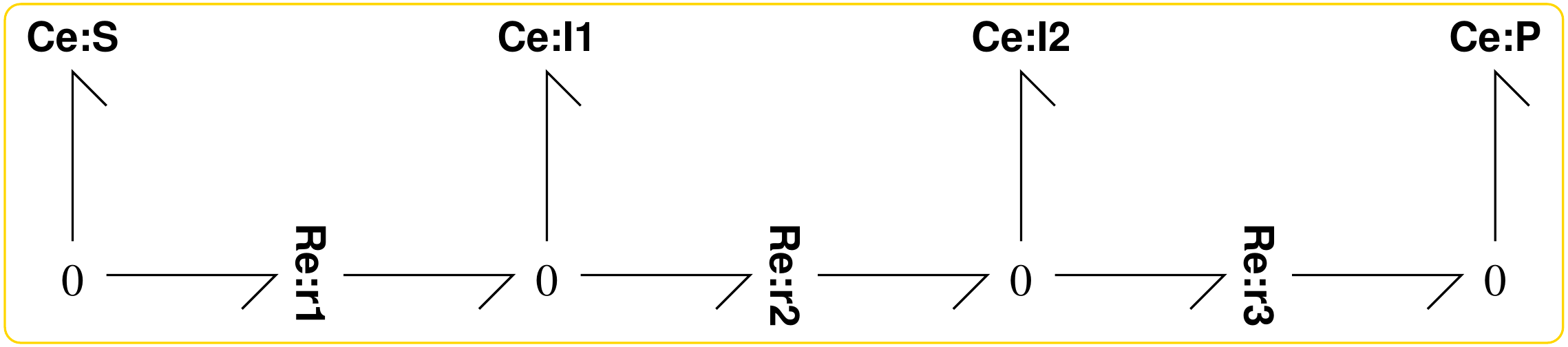}{Pathway (module \textbf{PATH})}{0.7}
  \caption{Enzyme-catalysed reaction (ECR).
    The Bond Graph  notation is: $\rightharpoondown$ energy
    connection; \Ce species; \Re reaction; \Zero~common potential
    connection; \One~common flow connection~\cite{GawCra14}. Modules
    containing a bond graph and/or sub-modules are indicated by the
    notation \textbf{module:instance}~\citep{GawCra16,Gaw17a};
    thus in (b) \textbf{ecr:ecr}
    represents an instance called ``ecr'' of the module ``ecr''
    containing the bond graph  of (a). Module \emph{ports} are denoted by
    []; thus port [A] of \textbf{ecr} corresponds to the component
    \BCe{A} of (a).
    (a) Enzyme-catalysed reaction with competitive
    activation and inhibition. \BCe{A}, \BCe{B}, \BCe{E}, \BCe{C},
    \BCe{Act} \& \BCe{Inh} represent the substrate, product, enzyme,
    enzyme-substrate complex, activation and inhibition respectively.
    In the sequel, the generic species such as \ch{A} will be replaced
    by specific species.
    \BCe{F} \& \BCe{G}, provide the driving energy. Example species
    appear in Figure \ref{fig:CFM}. 
    (b) A simple model of cooperativity is included by specifying that
    4 activation and 4 inhibition species interact with the enzyme;
    this is achieved in a modular fashion by incorporating the bond
    graph  of (a) into module \textbf{ecr}.
    (c) The module \textbf{SYS} (the controlled system) of
    Figure~\ref{subfig:fb_abg} is, in this example, a simple path of 3
    reactions represented by \BRe{r1}--\BR{r3} with intermediate
    species \BCe{I1} and \BCe{I2}.
  }
  \label{fig:ECR_abg}
\end{figure}
A network thermodynamics approach to modelling systems of biochemical
reactions was introduced by \citet{OstPerKat71,OstPerKat73} and
utilised the bond graph approach of \citet{Pay61}.
Tutorial introductions to bond graphs are available for control
engineers \cite{GawBev07} and systems biologists
\cite{GawPan20}. This section gives a brief introduction to bond
graphs focused on the applications in this paper.

Bond graphs focus on the energetic connection between components and
the $\rightharpoondown$
symbol indicates such an energetic connection; the
half-arrow indicates the direction corresponding to positive energy
flow.  In the biomolecular context, each such bond is associated with
two covariables: chemical potential~${\mu}~\Si{\joule\per\mole}$ and
flow $v~\Si{\mole\per\second}$. The key point is that the product of $\mu$
and $v$ is power $p=\mu v~\Si{\watt}$. Alternatively, it is possible to
scale these co-variable by Faraday's constant $F~\Si{\coulomb\per\mole}$
to give $\phi=\frac{1}{F}\mu~\Si{\volt}$ and $f=Fv~\Si{\ampere}$ where
\Si{\joule\per\coulomb} has been replaced by the more convenient unit volt
\Si{\volt} and \Si{\coulomb\per\second} has been replaced by the more
convenient unit ampere \Si{\ampere} \citep{Gaw17a}.

Electrical components may be connected in \emph{parallel} (where the 
\emph{voltage} is common) and \emph{series} (where the \emph{current}
is common). These two concepts are generalised in the bond graph
notation as the \zero junction which implies that all impinging bonds
have the same \emph{potential} (but different flows) and the
\one junction which implies that all impinging bonds have the same
\emph{flow} (but different potentials). The direction of
positive energy transmission is determined by the bond half arrow.
As all bonds impinging on a \zero junction have the same
\emph{potential}, the half arrow implies the sign of the
\emph{flows} for each impinging bond. The reverse is true for \one
junctions, where the half arrow implies the signs of the \emph{potentials}.

In this context, \emph{species} are represented \Ce components and
\emph{reactions} by \Re components.
In particular, consider the
three reactions
\begin{align*}
\ch{E + Inh &<>[ r0 ] Act + E0 }\\
\ch{A + E + F &<>[ r1 ] C }\\
\ch{C &<>[ r2 ] B + E + G }
\end{align*}
which are represented by the bond graph  of Figure~\ref{subfig:ecr_abg}.
The species \ch{A}, \ch{B}, \ch{C} etc. are represented by the
components \BCe{A}, \BCe{B}, \BCe{C} etc.; the reactions \ch{r1},
\ch{r2} and \ch{r3} are represented by the components \BRe{r1},
\BRe{r2} and \BRe{r3}.
Thus, for example, the lower left \one junction ensures that the flow
from the species \ch{A}, \ch{E} and \ch{F} is identical to that though
the reaction \ch{r1}. Energy conservation at the junction then implies
that the net potential impinging on the left of reaction \ch{r1} is
the sum of the species potentials.
Similarly, the lower centre \zero junction ensures that the potential
of species \ch{C} appears at the right of reaction \ch{r1} and the
left of reaction \ch{r2}. Energy conservation at the junction then
implies that the flow into species \ch{C} is the difference of the
flows though reactions \ch{r1} and \ch{r2}.

Figure~\ref{subfig:ecr_abg} clearly shows the structure of the
reaction network in terms of which species are connected to which
reactions.
%
%
As discussed in the caption, Figure \ref{subfig:ECR_abg} contains the
module \textbf{ecr:ecr} containing the bond graph of
Figure~\ref{subfig:ecr_abg}; and the four components \BCe{A}, \BCe{B},
\BCe{Act} \& \BCe{Inh} are exposed as ports.
Thus in Figure \ref{subfig:ECR_abg},
the arrangement of four bonds collecting \zero and \one junctions
changes the stoichiometry of the first reaction to:
\begin{align*}
\ch{E + 4 Inh  &<> [ r0 ] E0 + 4 Act   }
\end{align*}

For the purposes of this paper (see \cite{GawCra14} for more detail),
the mass-action kinetics of the reaction system corresponding to a bond graph
representation is obtained from the \Ce constitutive relation:
\begin{equation}\label{eq:CR_C}
  \phi_A = RT \ln K_A x_A
\end{equation}
where $\phi_A$ and $x_A$ are the chemical potential (\si{\joule\per\mole}) and concentration
relative to standard conditions (dimensionless) of the generic species
\ch{A}, $K_A$ is a dimensionless constant \cite[\S~1.2]{KeeSne09} and
$RT$  (\si{\joule\per\mole}) is the product of the universal gas constant $R$ and absolute
temperature $T$;
the  \BRe{ri} constitutive relation for the $i$th reaction 
\begin{equation}\label{eq:CR_R}
  v_i = \kappa_i \left [ \exp \frac{\Phi^f}{RT} -  \exp \frac{\Phi^r}{RT}  \right ]
\end{equation}
where $v_i$ is the ratio of the $i$th reaction flow to a nominal flow (dimensionless), $\kappa_i$
(dimensionless) is the reaction constant. $\Phi^f$ and
$\Phi^r$  (\si{\joule\per\mole}) are the forward and reverse reaction affinities of the ith
reaction determined by the species potentials $\phi$ and the structure
of bonds and junctions determining the stoichiometry of the reaction
networks. Different forms of the \BRe{ri} constitutive relation
(\ref{eq:CR_R}) give rise to different forms of kinetics \cite{GawCudCra20}.

Systems of chemical reactions written in terms of rate constants, and
the corresponding ODEs are not necessarily consistent with physical
principles \cite[\S~1.3]{KeeSne09}. In contrast, chemical reactions
written as bond graphs automatically obey the Wegscheider conditions
\cite{GawCurCra15}. Thus reaction systems modelled in bond graph form
can be combined in a modular way \cite{GawCra16}; the resultant
modular bond graph can then be used to generate and energetically
consistent set of ODEs suitable for simulation. Similarly, bond graphs
account for the phenomenon of \emph{retroactivity} \cite{VecNinSon08} in
a transparent fashion \cite{GawCra16}.

\section{Bond graph based control analysis}
\label{sec:bond-graph-based}
\begin{figure}[htbp]
  \centering
  \SubFig{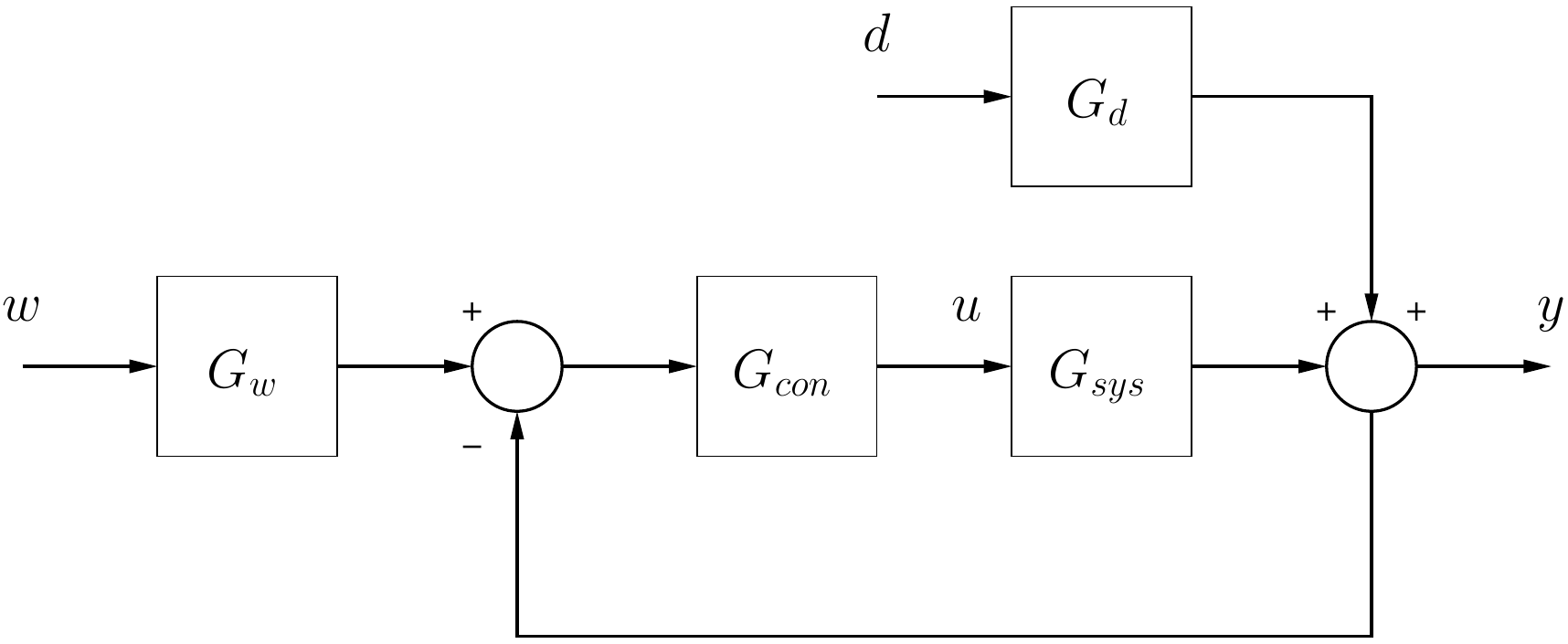}{Transfer function}{0.7}
  \SubFig{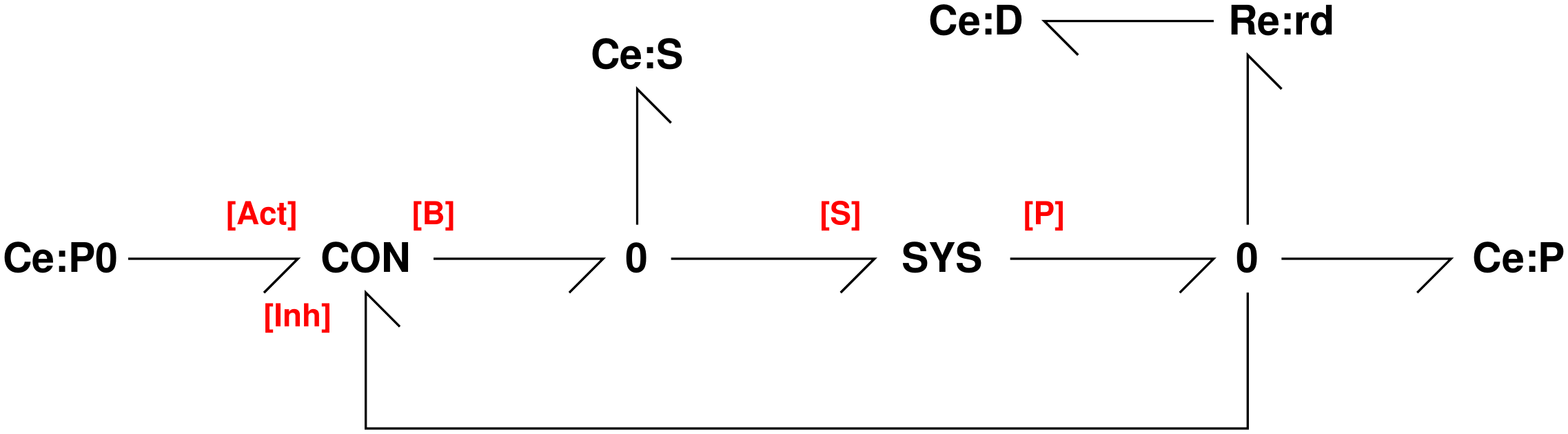}{Bond graph}{0.8}
  \caption{Feedback control.  %
    (a) A classical feedback loop block diagram representing the
    linearisation of a non-linear system. The blocks represent
    transfer functions which are connected by signals. $y$ is the
    controlled output, $d$ a disturbance and $w$ the setpoint, or
    desired value of $y$.  %
    (b) A bond graph feedback loop. \textbf{CON} and \textbf{SYS} are
    bond graph \emph{modules} with \emph{ports} denoted by
    [].
    The module \textbf{CON} will be replaced by various modules in the
    sequel. In particular, module \textbf{ECR} of
    Figure \ref{subfig:ECR_abg} is used as \textbf{CON} with three
    components exposed; the hidden component \BCe{A} is a chemostat in
    the sequel.
    \BCe{P}, \BCe{D} and \BCe{P0} are bond graph components
    representing species corresponding to product, disturbance and
    reference species respectively.
    Throughout this paper, the \textbf{SYS} component is the pathway
    system \textbf{Path} of Figure \ref{subfig:Path_abg}.
    The $\rightharpoondown$ symbol indicates an energetic connection
    between two subsystems; the half-arrow indicates the direction
    corresponding to positive energy flow. 
    The relationship between the block diagram of (a) and the bond
    graph of (b) is examined in \S~\ref{sec:open-loop}.  }
  \label{fig:feedback}
\end{figure}
Figure \ref{subfig:feedback} depicts a conventional feedback control
system in transfer-function form.
The four transfer functions $G_{con}(s)$, $G_{sys}(s)$, $G_{w}(s)$ and $G_{g}(s)$
represent the controller, the system under control, the setpoint and
disturbance transfer functions respectively where $s$ is the Laplace variable.
The four signals $y$, $u$, $w$ and $d$ represent the system output,
system input, setpoint and disturbance respectively. In this context,
$y$ could be a product whose concentration is to be controlled, $u$ a
reaction flow, $w$ the ideal product concentration and $d$ the
concentration of another species.

The closed-loop transfer function is:
\begin{align}
  y &= \frac{\Ls}{1+\Ls}\Gs{w}w +
      \frac{1}{1+\Ls}\Gs{d}d   \label{eq:cl} \\
  \text{where } \Ls &= G_{con}(s)G_{sys}(s)\label{eq:L(s)}
\end{align}
$\Ls$ is referred to as the feedback \emph{loop gain}.
In the engineering context, $G_{con}(s)$ and $G_{sys}(s)$ would arise
from separate physical entities; nevertheless, the loop gain $\Ls$
(\ref{eq:L(s)}) appearing in equation (\ref{eq:cl}) only requires the
\emph{product} of $G_{con}(s)$ and $G_{sys}(s)$. This is important for
biomolecular systems where there is no clear physical distinction
between controller and system: it is the feedback loop itself that is
of fundamental importance.

Typically, such control systems are analysed in the frequency domain
by setting $s=j\omega$ where $j=\sqrt{-1}$ and $\omega$ is frequency
in \si{\radian\per\sec}. At those frequencies where $\Lw$ is
large,
equation (\ref{eq:cl}) can be approximated by
  $y \approx \Gs{w}w$.
In other words, a large loop gain $\Lw$ is desirable insofar as the
system output $y$ is a close match to the desired value $ \Gs{w}w$
despite disturbances represented by $d$. However, incorrect choice of
the the loop gain $\Ls$ can lead to instability and $\Ls$ is, moreover, 
subject to fundamental constraints \cite{AstMur08}.

To summarise, there are two potentially conflicting issues in
controller design: good disturbance rejection and stability; these are
both captured in the loop gain $\Ls$.

Figure \ref{subfig:feedback} implicitly assumes that the connection
between subsystems, such as those represented by $G_{con}(s)$ and
$G_{sys}(s)$ is one-way as indicated by the arrows. However, the
physical controller needs to be designed to make sure this one-way
interaction is correct; this requires the use of energy.
It has been argued that this approach is misguided, even in the
context of engineering systems. This has lead to the concept of
physical-model based control
\cite{Kar79a,ShaHogHar91,Gaw95c}

In the context of biomolecular systems, the concept of
\emph{retroactivity} \citep{VecNinSon08} has been introduced
to explain why interaction is not one-way and thus design based on
simplistic application of the approach of Figure \ref{subfig:feedback}
often fails.

There are two reasons why the bond graph approach is superior to the
block diagram approach of Figure \ref{subfig:feedback} in the
context of feedback control:
\begin{enumerate}
\item It explicitly accounts for the two-way interaction found in
  physical systems in general and biomolecular systems in particular.
\item It explicitly accounts for energy flows and thus can directly
  expose performance/energy consumption trade-offs; this is the
  subject of current research.
\end{enumerate}
For this reason, the transfer function paradigm of Figure
\ref{subfig:feedback} is replaced by the bond graph based paradigm of
Figure \ref{subfig:fb_abg}.

Figure \ref{subfig:fb_abg} is based on the notation for modular bond
graphs \citep{GawCurCra15}.
The two bond graph modules are \textbf{CON} and \textbf{SYS};
\textbf{CON} represents the controller and has three ports: [Act]
(activation), [Inh] (inhibition) and [Con] (control signal) and
\textbf{SYS} represents the system and has two ports: [S] (substrate)
and [P] product. In the sequel, the system module \textbf{SYS} is the
pathway \textbf{Path} module of Figure \ref{subfig:Path_abg} but the controller module will be
instantiated by three modules in turn: an enzyme catalysed reaction
with competitive activation and inhibition
(\S~\ref{sec:exampl-enzyme-cataly}), cyclic flow modulation
(\S~\ref{sec:CFM}) and cyclic flow modulation with integral action
(\S~\ref{sec:feedback-control}).

The components \textbf{Ce:P} and \textbf{Ce:S} represent the product
and substrate species respectively and the components \textbf{Ce:P0}
and \textbf{Ce:D} represent the reference species and product
disturbance respectively; because \textbf{Ce:P0} and \textbf{Ce:D}
represent exogenous variables, they are \emph{chemostats}
\citep{PolEsp14,GawCra16}.

Because the paper combines approaches from three fields -- systems
biology, bond graphs and control theory -- there are three notational
systems. To summarise, a generic species \ch{A} has a bond graph
representation \BCe{A} and the corresponding ``signal'' is the
concentration $x_A$. In a particular case, the generic species \ch{A}
can be instantiated as a particular species such as
fructose-6-phosphate \Ch{F6P}.

As shown in the sequel, the bond graph modelling approach can
make use of the transfer function approach to understand the dynamic
properties of feedback systems of the form of Figure
\ref{subfig:fb_abg}. In particular, as shown in \S~\ref{sec:open-loop},
the fundamental control systems concept of loop-gain can be retrieved
from the bond graph modelling paradigm. But first, linearisation must
be considered.


\subsection{Linearisation}
\label{sec:linearisation}
Biomolecular systems are nonlinear and must be linearised before
applying transfer function techniques. Linearisation of biomolecular
systems in a biomolecular context, together with a discussion on
retroactivity, is given by \cite{GawCra16}. In particular, the
non-linear system equations are:
\begin{xalignat}{2}
  \ddt{\xx}& = \NN\ff&
  \ff &= F(\xx,\xxc)
\end{xalignat}
In systems biology terms: the $n_\xx$ vector $\xx$ represents the
amount of each non-chemostatted species (\si{\mole}), the
$n_\xx \times n_\ff$ matrix $N$ is the system \emph{stoichiometric
  matrix}, the $n_\ff$ vector $\ff$ represents the flow in each
reaction (\si{\mole\per\second}).The $n_{\xxc}$ vector $\xxc$ represents
the amount of each chemostatted species (\si{\mole}).
$ F(\xx,\xxc)$ is a \emph{nonlinear} function of both arguments.
Because of thermodynamic constraints, $F$ has a particular structure
dependent on the stoichiometric matrix
$N$~\cite{GawCra18a} and is automatically generated from
the bond graph representation.
%
In standard control system terms, $\xx$ is the system \emph{state},
$\ff$ is the system \emph{output} and $\xxc$ the system \emph{input}.

The corresponding linearised equations are:
\begin{align}
  \ddt{\xt}& = \NN\ft\\
  \ft &= \Css \xt + \Dss \xtc\\
  \where
  \xt &= \xx - \xb\\
  \xtc &= \xxc - \xbc\\
  \and
  \ft &= \ff - \fb
\end{align}
where the $n_\ff \times n_\xx$ matrix $\Css$ and the
$n_\ff \times n_{\xxc}$ matrix $\Dss$ are given by the partial
derivatives:
\begin{xalignat}{2}
  \Css & = \frac{\partial \ff}{\partial \xx} &
  \Dss & = \frac{\partial \ff}{\partial \xxc} 
\end{xalignat}
evaluated at the steady-state values $\xb$ and $\fb$ of state and flow
respectively corresponding to the constant chemostat state
$\xxc=\xbc$:
\begin{xalignat}{2}
  \NN\fb & = 0&
  \fb &= F(\xb,\xbc)
\end{xalignat}
Linearisation has two steps: finding the steady-state state $\xb$ and
flow $\fb$
and then computing the linearisation matrices $\Css$ and $\Dss$.
The first is simply accomplished by numerically simulating the system
until a steady-state is reached ($\ddt{\xx}\approx 0$). The second is
achieved symbolically within \texttt{BondGraphTools}
(\url{https://pypi.org/project/BondGraphTools}) using the symbolic
derivative functions of the sympy library
(\url{https://www.sympy.org}).
The Python Control Systems Library
(\url{https://pypi.org/project/control/}) is used to convert the
linearised system from state-space form to transfer function form,
manipulate transfer functions and to generate time and frequency
responses.

\subsection{Example: Enzyme-catalysed reaction control}
\label{sec:exampl-enzyme-cataly}


\begin{figure}[htbp]
  \centering
  \Fig{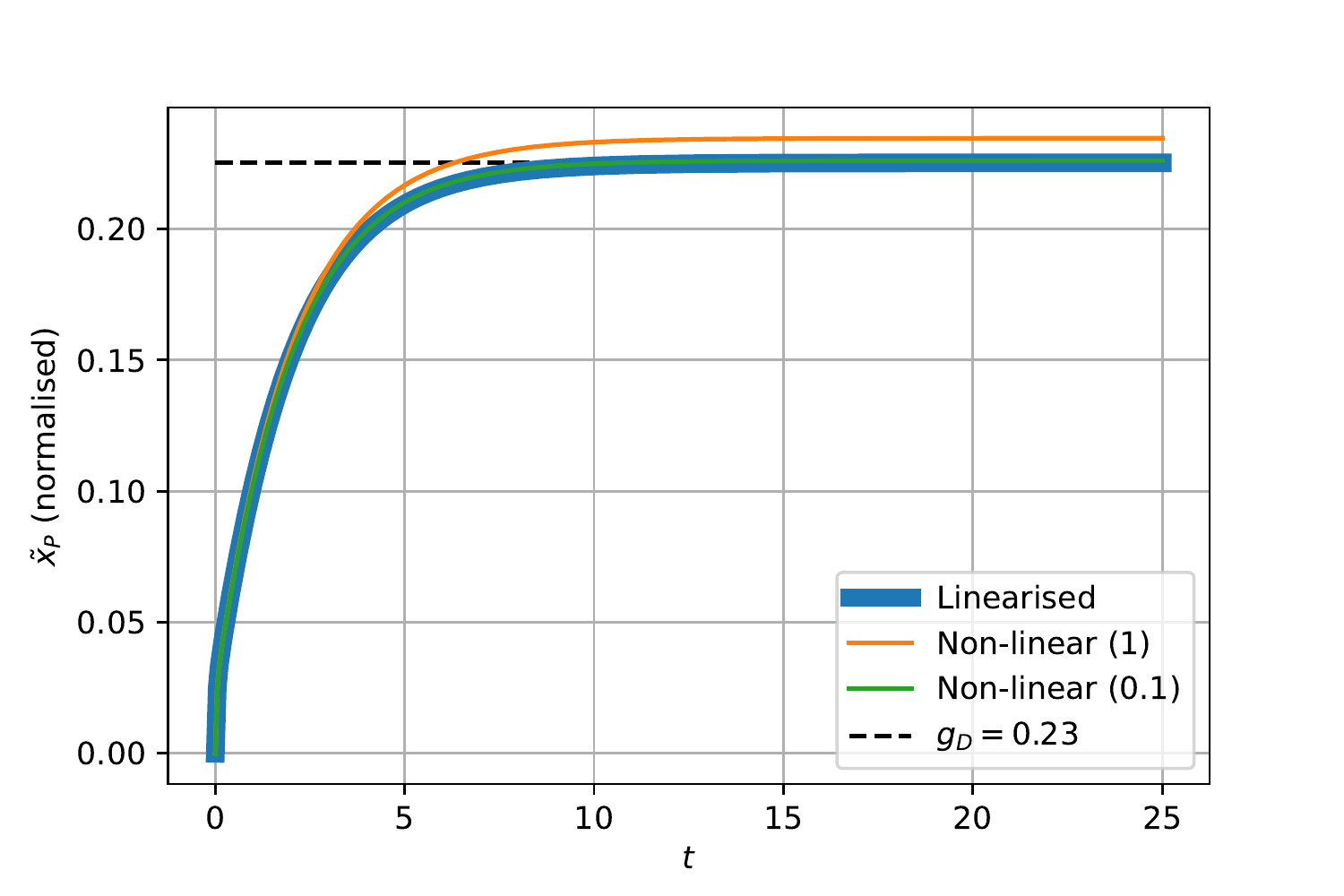}{0.6}
  \caption{Non-linear and linearised closed-loop step response. The
    asymptotic value $g_D$ is indicated by a dashed line. The
    amplitude of the disturbance  step is given for the non-linear simulations and
    the resultant response is divided by the step amplitude. The
    normalised nonlinear response is close to the linear case for an
    amplitude of 0.1 and differs slightly for an amplitude of 1.0.
    For the purposes of illustration, all parameters (which are
    dimensionless - see \S~\ref{sec:bond-graph-based-1}) are unity
    except for the species constants for \ch{F} and \ch{G}:
    $K_F=10^3$,$K_G=10^{-3}$ and reaction constants for the controller
    (Fig. \ref{subfig:ecr_abg}):
    $\kappa_{r1}=\kappa_{r2} = 2$ and system
    (Fig. \ref{subfig:Path_abg}): $\kappa_{r1}=\kappa_{r2} = \kappa_{r3} =
    10$
    The steady-state values  (which are also
    dimensionless - see \S~\ref{sec:bond-graph-based-1})  were $x_S = 12.85$ and $x_P = 10.12$.
  }
  \label{fig:step}
\end{figure}

The modified enzyme-catalysed reaction module of Figure
\ref{subfig:ECR_abg} and the pathway module of Figure
\ref{subfig:Path_abg} are embedded in the feedback loop of Figure
\ref{subfig:fb_abg} as \textbf{CON} and \textbf{SYS} respectively and
used for the purposes of illustration; the parameters are given in
Figure \ref{fig:step}.

The non-linear system equations were derived from the modular bond
graph of Figure \ref{subfig:fb_abg} using \texttt{BondGraphTools} and
simulated to give the steady-state condition corresponding to the
parameters of Figure \ref{fig:step}. The linearised equations were
then extracted and the transfer function relating the disturbance
$\xt_D$ to the product $\xt_P$ generated. The corresponding
closed-loop step response appears in Figure \ref{fig:step}.

However, simulation does not provide an explanation of why the
steady-state is the particular value shown nor why the dynamics are as
shown.  The explanation is provided by the analysis of the following
section.



\subsection{Open-loop analysis}
\label{sec:open-loop}

\begin{figure}[htbp]
  \centering
  \SubFig{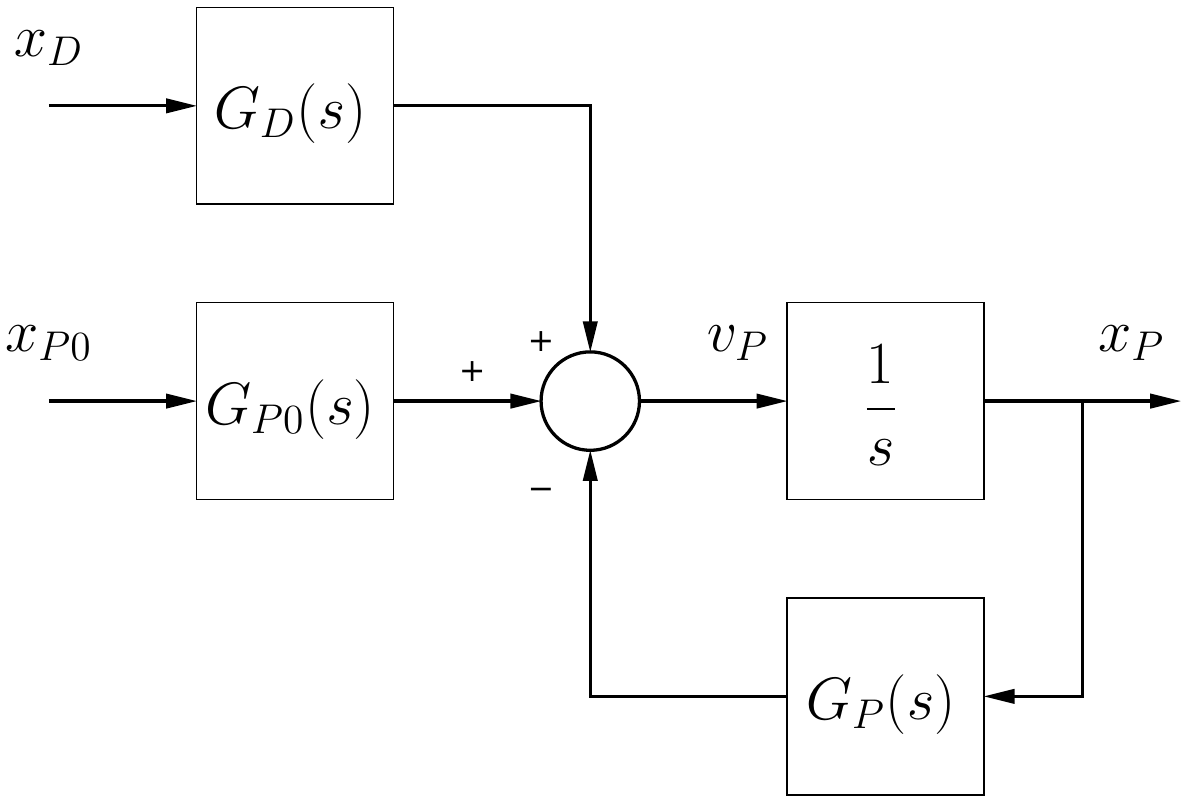}{Loop analysis}{0.45}
  \SubFig{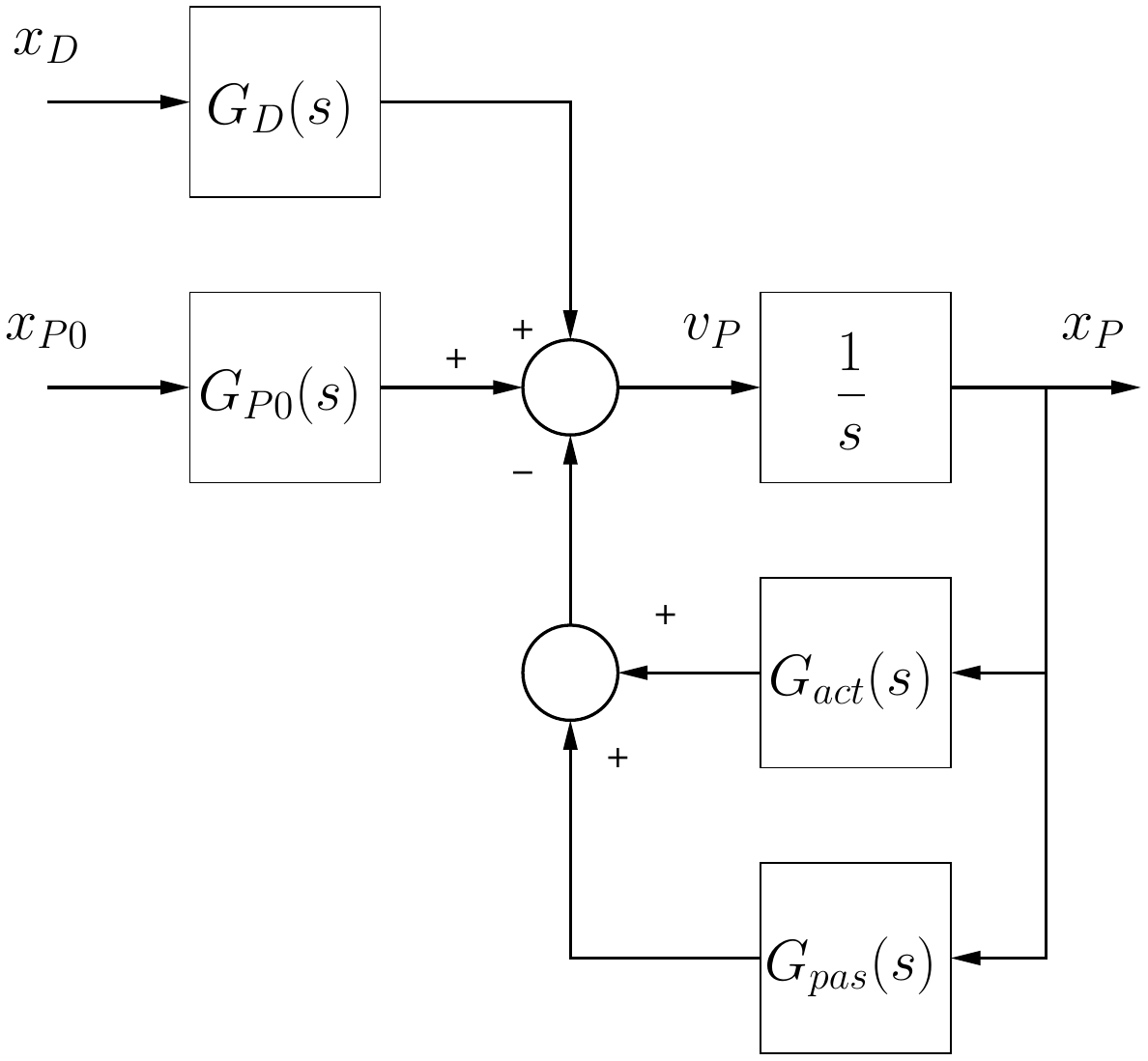}{Split-loop analysis}{0.45}
  \caption{Loop analysis.
    (a) The block diagram corresponding to opening the bond graph
    feedback loop by setting the product \BCe{P} to be a
    chemostat. $x_P$ is the amount of product and $v_P$ the product
    flow. $\Gs{P}$, $\Gs{D}$ and $\Gs{pP0}$ are transfer functions
    relating $\xx_{P}$, $\xx_{D}$ and $\xx_{P0}$ to $\ff_{P}$.
    (b) The transfer function $\Gs{P}$ is split into two terms:
    $\Gs{act}$ and $\Gs{pas}$ corresponding to active and passive feedback.
  }
  \label{fig:bgfb_ol}
\end{figure}

As discussed above, the loop-gain $\Ls$ is a key transfer function in
the classical control systems analysis of Figure
\ref{subfig:feedback}. This section indicates how the loop-gain $\Ls$
can be derived from the bond graph of Figure \ref{subfig:fb_abg}.

The closed-loop system of Figure \ref{subfig:fb_abg} includes two
chemostats \BCe{P0} and \BCe{D} which make the corresponding states
$x_{P0}$ and $x_D$ independent variables; the product state $x_P$
remains a dependent variable which evolves with time as in Figure
\ref{fig:step}. To create an open loop system, the component \BCe{P}
representing the product is also made a chemostat thus making $x_P$ an
independent variable.

The linearised flow $\ft_P$ into the chemostat \BCe{P} is given by the sum of
three terms corresponding to the three chemostats \BCe{P}, \BCe{P0} and
\BCe{D} respectively:
\begin{equation}\label{eq:vt_p}
  \ft_P = -G_{P}(s) \xt_P + G_{P0}(s) \xt_{P0} + G_D(s) \xt_D
\end{equation}
where $-G_{P}$, $G_{P0}$ and $G_{D}$ are the transfer functions
relating $\ft_P$ to $\xt_P$, $\xt_{P0}$ and $\xt_D$ respectively. The
minus sign associated with $G_{P}$ is to give compatibility with
standard definitions of loop gain in a negative feedback context.

To reclose the loop, \BCe{P} is restored to non-chemostatted dynamics using
the transfer function relating $\xt_P$ to $\ft_P$:
\begin{equation}\label{eq:xt_p}
  \xt_P = \frac{1}{s} \ft_P
\end{equation}
The block diagram corresponding to Equations (\ref{eq:vt_p}) and
(\ref{eq:xt_p}) is shown in Figure \ref{subfig:loop}. Using Equation
(\ref{eq:L(s)}), the loop gain $\Ls$ is given by:
\begin{equation}
  \Ls = \frac{G_p(s)}{s}
\end{equation}

From the block diagram of Figure \ref{subfig:loop}, or from Equations
(\ref{eq:vt_p}) and (\ref{eq:xt_p}), the closed-loop system can be
explicitly written as
\begin{equation}\label{eq:xt_P}
  \xt_P = \frac{1}{s+G_P(s)} \left [ G_{P0}(s) \xt_{P0} + G_D(s) \xt_D \right ]
\end{equation}

The steady state value $\xtb_P$ of $\xt_p$ is obtained by setting
$s=0$ to give
\begin{equation}
  \xtb_P = \frac{1}{G_P(0)} \left [ G_{P0}(0) \xtb_{P0} + G_D(0) \xtb_D \right ]
\end{equation}
In particular, the steady-state disturbance gain $g_D$ is given by:
\begin{equation}\label{eq:g_D}
  g_D = \frac{\xtb_P}{\xtb_D} = \frac{G_D(0)}{G_P(0)}
\end{equation}
With the parameters given in Figure \ref{fig:step}:
\begin{align}
  G_D(0) &= 1.00\\
  G_P(0) & = 4.44\\
  \hence
  g_D& = 0.23\label{eq:g_D_ECR}
\end{align}
This corresponds to Figure \ref{fig:step}.



\subsection{Split Loop analysis}
\label{sec:split-loop-analysis}
\begin{figure}[htbp]
  \centering
  \Fig{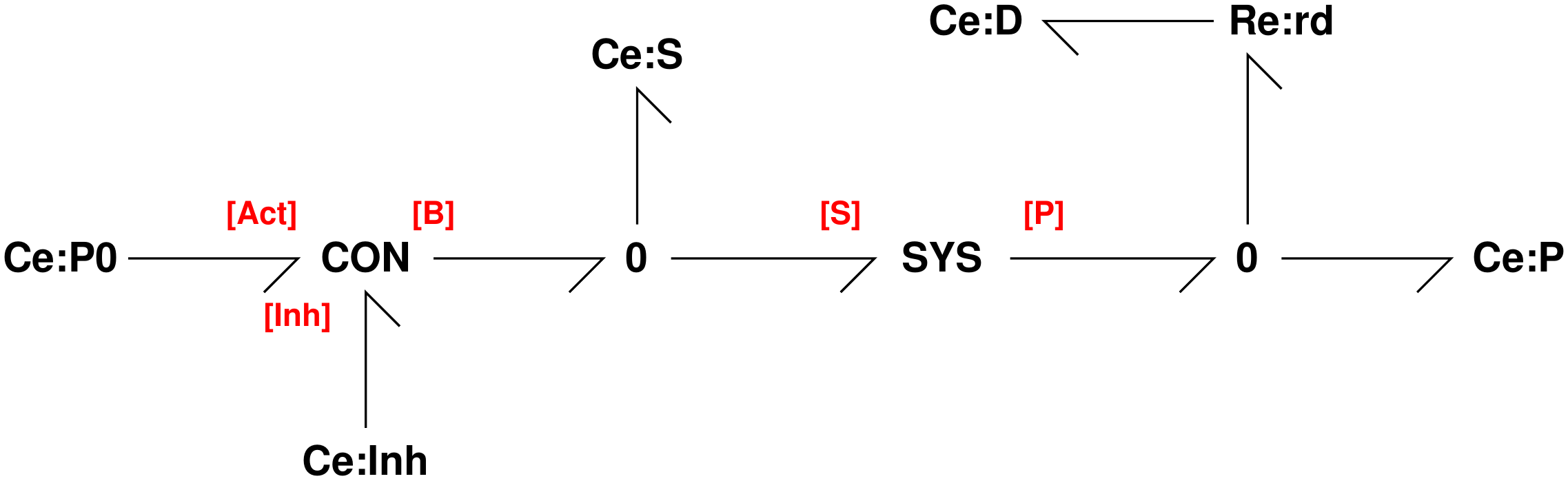}{0.9}
  \caption{Split-loop.
    The feedback bond in Figure \ref{subfig:fb_abg} is removed and
    replaced by the chemostat \BCe{Inh}. This splits the loop and
    allows active and passive feedback to be distinguished as
    described in the text.
  }
  \label{fig:split}
\end{figure}
The previous section shows how the loop gain $\Ls$ may be derived
from the closed-loop system in the bond graph  form of Figure
\ref{subfig:fb_abg}.
This section expands this analysis by dividing the loop gain $\Ls$
 into two parts: and \emph{active} part $\Las$
and a passive part $\Lps$ so that
\begin{equation}\label{eq:L_split}
  \Ls = \Las + \Lps
\end{equation}
The active part arises mainly from the properties of the controller (\textbf{CON}); the
passive part arises mainly from the properties of the system
(\textbf{SYS}) appearing in the closed-loop bond graph of Figure
\ref{subfig:fb_abg}.

The \emph{split-loop} procedure is based on removing the feedback bond
linking the controlled product \BCe{P}
to the inhibition port ([Inh]) of the controller. This is depicted in
Figure \ref{fig:split} where the bond has been removed and the
chemostat \BCe{Inh} has been added.
To focus on the loop gain, the chemostats \BCe{P0} and
\BCe{D} are held at the steady state values ($\xt_D = \xt_{P0}=0$) for
the rest of this section.
The linearised flows $\ft_{PP}$ into the chemostat \BCe{P} and
$\ft_{II}$ into the chemostat \BCe{Inh} are each given by the
sum of two terms corresponding to the two variable chemostats
\BCe{Inh} and \BCe{P} respectively:
\begin{align}\label{eq:vt_p_split}
  \ft_{PP} &= -G_{PI}(s) \xt_{Inh} - G_{PP}(s) \xt_P\\
  \ft_{II} &= -G_{II}(s) \xt_{Inh} - G_{IP}(s) \xt_P
\end{align}
When the split-loop is reconnected
\begin{align}
  \xt_{Inh} &=  \xt_P\\
  \text{and }
  \ft_P &= \ft_{PP} + \ft_{II}\\
            &= -\left [ G_{PI}(s) +  G_{PP}(s) +
              G_{II}(s) + G_{IP}(s) \right ] \xt_P
\end{align}
$G_{PI}(s)$ is the transfer function from the inhibition port of the
controller to the product and is thus the active part of the
control. Hence the previous equation is rewritten as:
\begin{align}
  \ft_P &= -\left [ G_{act} (s)  + G_{pas} (s) \right ] \xt_P\\
  \where
  G_{act} (s)  &=  G_{PI}(s) \\
  \and
   G_{pas} (s)&=  G_{PP}(s) +
              G_{II}(s) + G_{IP}(s) 
\end{align}

Once again, the minus signs associated with $G_{act}$ and $G_{pas}$
are to give compatibility with standard definitions of loop gain in a
negative feedback context.

To allow comparison with Equation (\ref{eq:vt_p}), the transfer
functions appearing Equation (\ref{eq:vt_p_split}) are evaluated with
the \emph{same steady states as those of the closed-loop system} and,
in addition, reconnection of the split loop implies
\begin{xalignat}{2}
  \xb_{inh} &= \xb_P&
  \xt_{inh} &= \xt_P
\end{xalignat}
Comparing  Equations (\ref{eq:vt_p} and Equation
(\ref{eq:vt_p_split}), it follows that:
\begin{equation}\label{eq:G_split}
  G_P = G_{act} + G_{pas}
\end{equation}
Further, defining
\begin{xalignat}{2}
  \Las &= \frac{G_{act}(s)}{s} & \Lps &= \frac{G_{pas}(s)}{s}
\end{xalignat}
Equation (\ref{eq:L_split}) follows from Equation (\ref{eq:G_split}).
Thus the block-diagram of Figure \ref{subfig:loop} can be expanded to
give the block-diagram of Figure \ref{subfig:sloop}.

The conventional approach to feedback control in the engineering
context would regard $L_{pas}$ as an unwanted artefact to be
eliminated by correct design; similarly, in the life-sciences context,
$L_{pas}$ would be regarded as due to retroactivity and therefore
undesirable \citep{VecNinSon08}. A theme of this paper is that both these
attitudes are inappropriate in this context. In the engineering context, using such
interactions to improve control are well established as physical-model
based control
\cite{Kar79a,ShaHogHar91,Gaw95c}
In the systems biology context, this paper will show that $L_{pas}$
has a stabilising influence on the control system.

\begin{figure}[htbp]
  \centering
  \SubFig{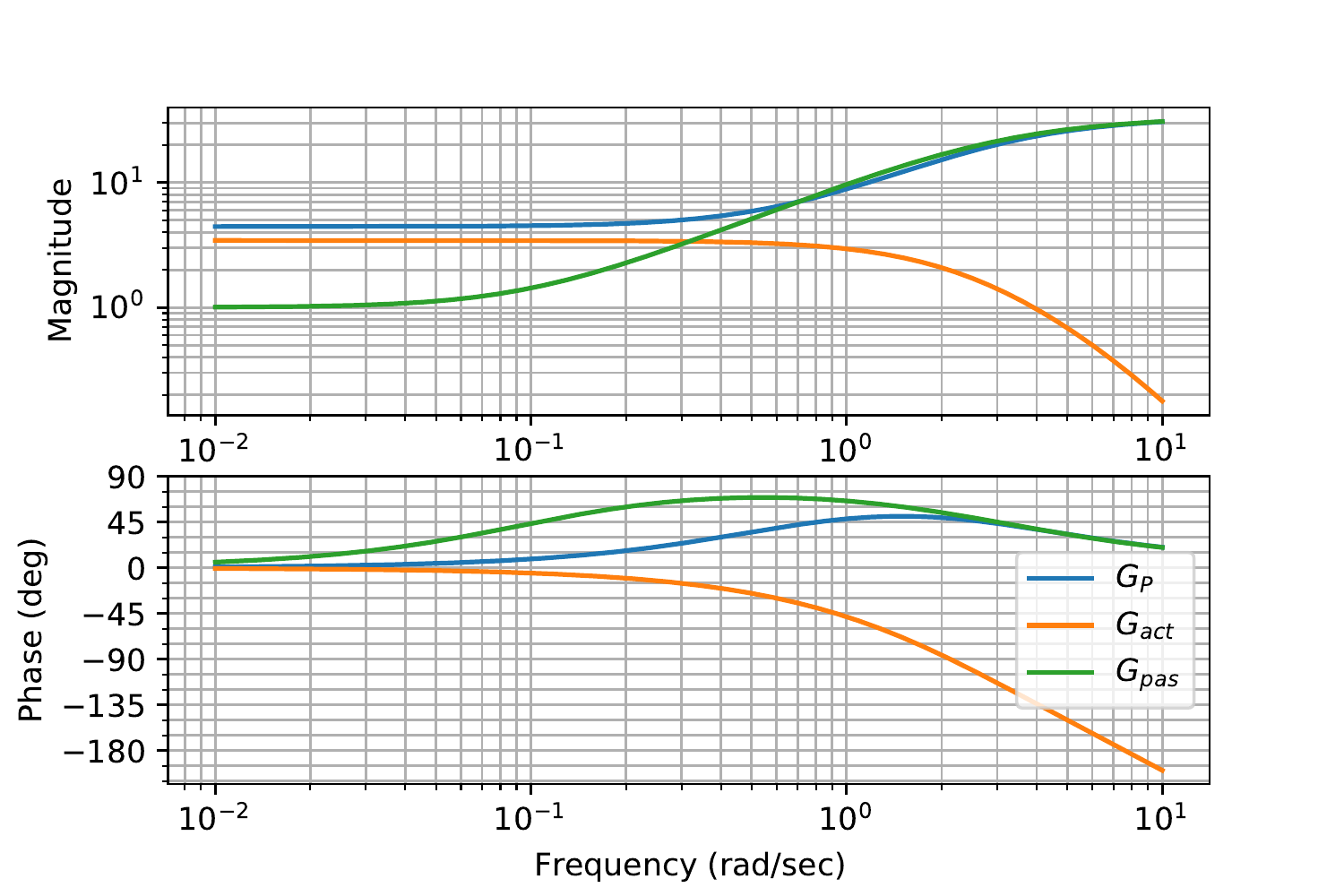}{Components of $G_P(s)$}{0.9}
  \SubFig{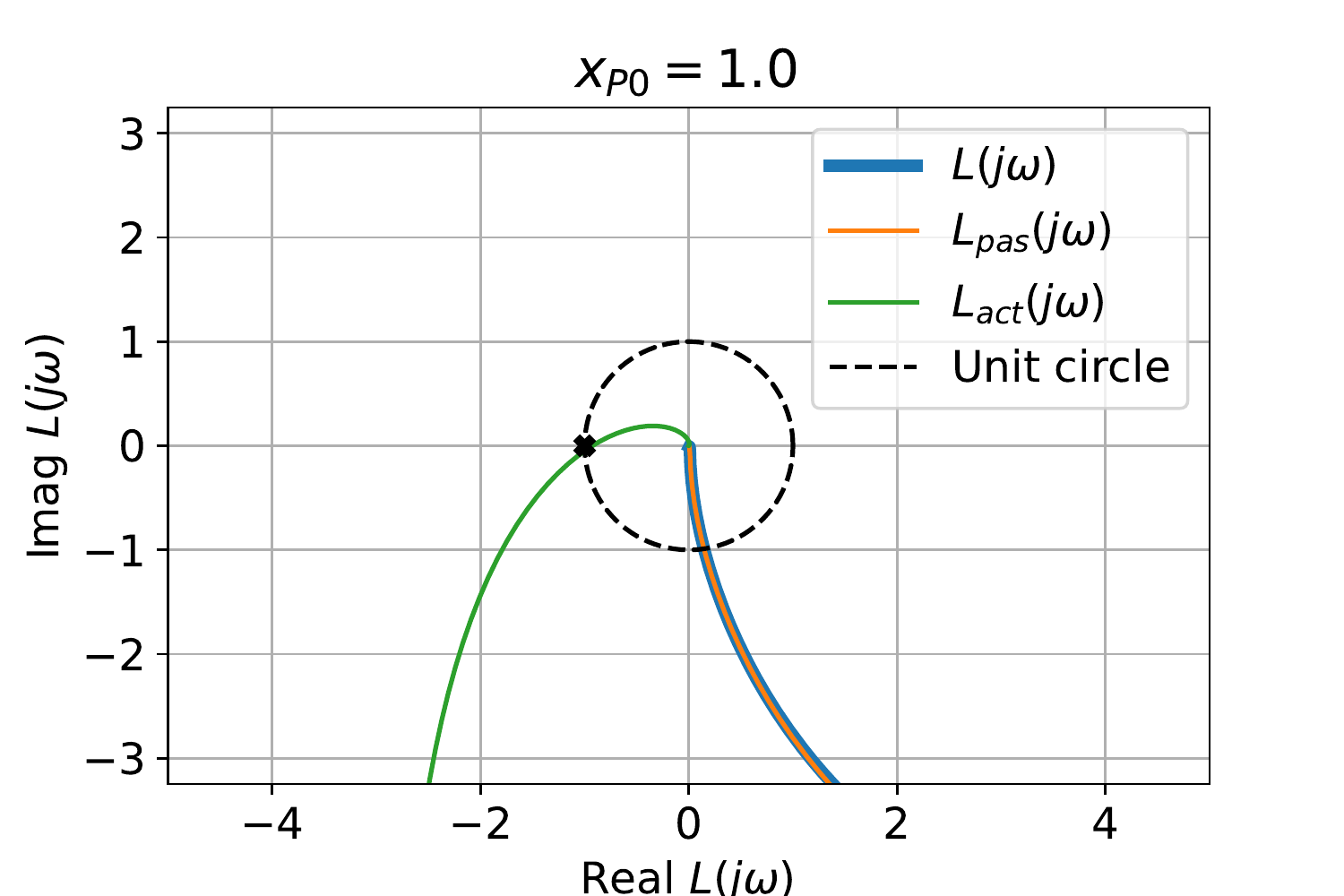}{Open-loop frequency response}{0.45}
  \SubFig{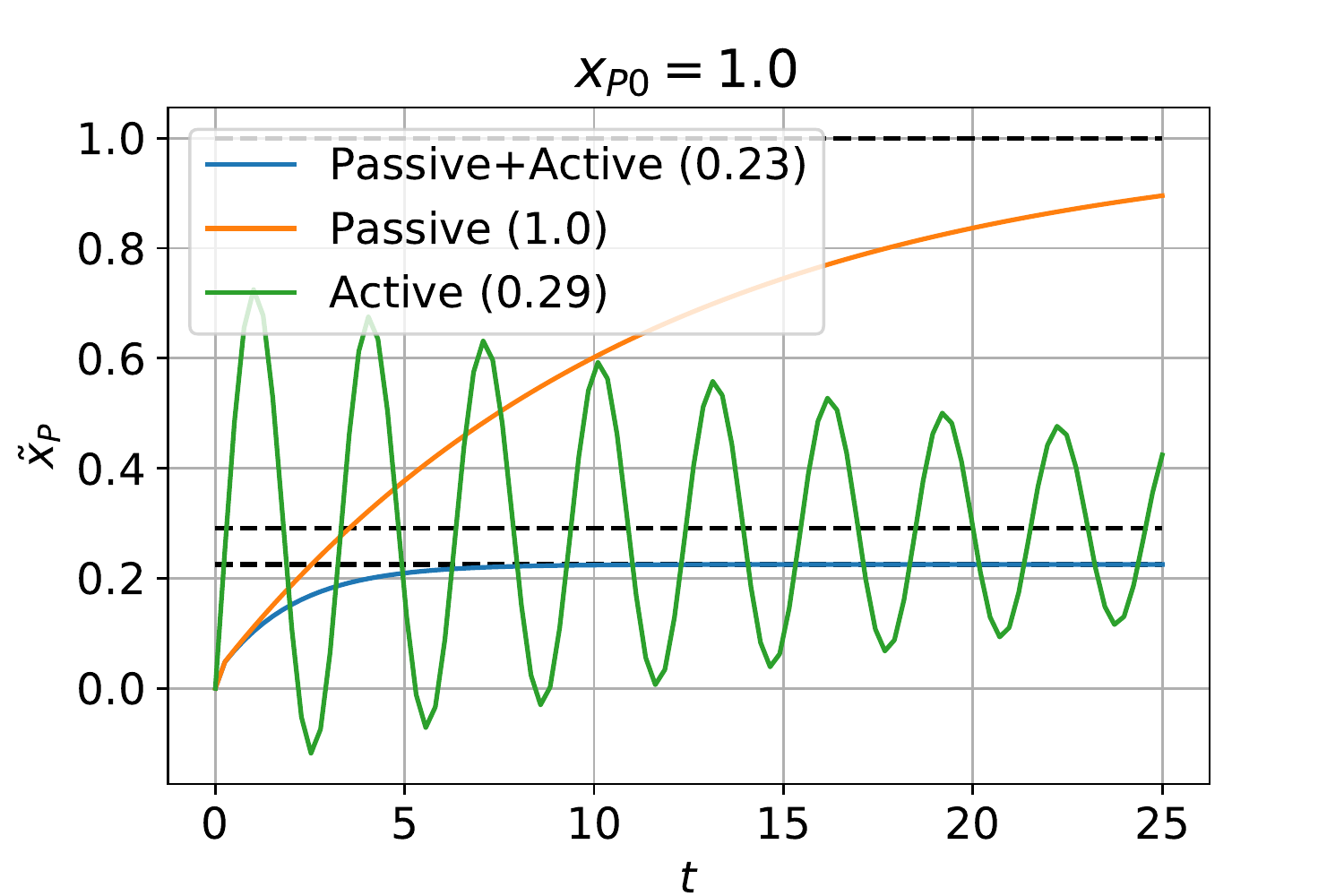}{Closed-loop step response}{0.45}
  \caption{ECR: Split-loop analysis.
    (a) Components of $G_P(\jw)$. $G_{PI}\approx 0$ and is not shown. Of the
    two remaining components of $G_{pas}$, $G_{II}$ is small at low
    frequencies and so $G_{pas}\approx G_{PP}$ at low frequencies; at
    mid and high frequencies, $G_{pas}$ provides phase advance.
    $G_{act}$ is large at low frequencies and small at high
    frequencies. Thus $G_P(\jw)\approx G_{act}(\jw)+G_{PP}(\jw)$ at low
    frequencies and $G_P(\jw)\approx G_{pas}(\jw)$ at high frequency.
    Thus $G_{act}(\jw)$ provides high gain (and thus low steady-state
    error) at low frequencies and $G_{pas}(\jw)$ provides stabilising
    phase advance at mid frequencies.
    (b) Open-loop frequency response. The loop gain
    $L(\jw)=\frac{G_P(\jw)}{s}$ and its two components $L_{act}(\jw)$
    and $L_{pas}(\jw)$ are plotted on a Nyquist diagram. $L_{act}(\jw)$
    passes close to the $-1$ point and thus, without the term
    $L_{pas}(\jw)$, $L(\jw)$ would correspond to closed-loop system
    close to instability. However, at the relevant frequencies,
    $L(\jw)\approx L_{pas}(\jw)$ and is well away from the $-1$ point
    and thus corresponds to a stable closed-loop system.
    (c) Closed-loop disturbance step response. The hypothetical
    closed-loop responses corresponding to $L_{pas}(\jw)$ and
    $L_{act}(\jw)$ are well damped with large steady-state error and
    oscillatory with small steady-state error respectively. The
    actual response corresponding to $L(\jw)$ combines the best of both.
    }
  \label{fig:ECR_analysis}
\end{figure}
The closed-loop system is given in terms of $G_P(s)$ by Equation
(\ref{eq:xt_P}).
Because of the decomposition (\ref{eq:G_split}), it is possible to see
how the control system would, in principle, behave with only passive or
only active control. In particular, if $\xt_{pas}$ and
$\xt_{act}$ are the product concentration deviations in the two cases:
\begin{align}
  \xt_{pas} &= \frac{1}{s+G_{pas}(s)}
               \left [ G_{P0}(s) \xt_{P0} + G_D(s) \xt_D \right ]
               \label{eq:xt_pass}\\
  \xt_{act} &= \frac{1}{s+G_{act}(s)}
               \left [ G_{P0}(s) \xt_{P0} + G_D(s) \xt_D \right ]
               \label{eq:xt_act}
\end{align}
Figure \ref{subfig:bgsl_10_bodeG} shows $G_P(\jw)$, $G_{pas}(\jw)$ and
$G_{act}(\jw)$ plotted on a Bode diagram \citep{AstMur08}. The
magnitude of $G_{act}(\jw)$ is large at low frequencies and small at
high frequencies whereas the magnitude of $G_{pas}(\jw)$ is small at
low frequencies and high at high frequencies. Hence the $G_P(\jw)$ is
close to $G_{pas}(\jw)$ at high frequencies yet retains the high gain
at low frequencies due to $G_{act}(\jw)$.

In classical control theory, the \emph{frequency response} of the loop
gain reveals dynamical properties -- including stability -- of the
closed-loop system.
One such frequency-based approach is based on the Nyquist diagram 
\citep{AstMur08} where the imaginary part of $L(\jw)$ is plotted
against the real part of $L(\jw)$ for a range of frequencies. The
phase $\angle{L(\jw)}$ when the modulus $|{L(\jw)}|=1$ is of interest,
hence the \emph{unit circle} is plotted on the Nyquist diagram of
Figure \ref{subfig:bg_10_nyqL}.
There are three frequency responses plotted:
$L_{pas}(\jw)$ shows that, as the frequency response is well away from the $-1$
point, the time  response is well-damped;
$L_{act}(\jw)$ shows that, as the frequency response passes close to the $-1$
point, the time response is oscillatory;
$L(\jw)$ is similar to $L_{pas}(\jw)$ near the unit circle and
therefore also has a well-damped response.

The corresponding unit step responses appear in
Figure \ref{subfig:bg_10_allstep} along with the step response of
$\xt_P$ corresponding to Equation (\ref{eq:xt_P}). The disturbance response of the
passive-only system is well-behaved but the steady-state value is
large; in contrast, the disturbance response of the
active-only system is oscillatory but the steady-state value is
small. The overall controller combines the best of both responses: it
is well behaved with a small steady-state value.
The numerical steady-state values for the overall controller are given
in Equation (\ref{eq:g_D_ECR}); in a similar fashion:
\begin{align}
  g_{pas} &= \frac{1}{G_{pas}(0)} = 1/1.0 = 1\\
  g_{act} &= \frac{1}{G_{act}(0)} = 1/3.44 = 0.29
\end{align}
Thus the small steady-state value is largely due to the active part
of the control.


\section{Cyclic flow modulation (CFM)}
\label{sec:CFM}

\begin{figure}[htbp]
  \centering
  \SubFig{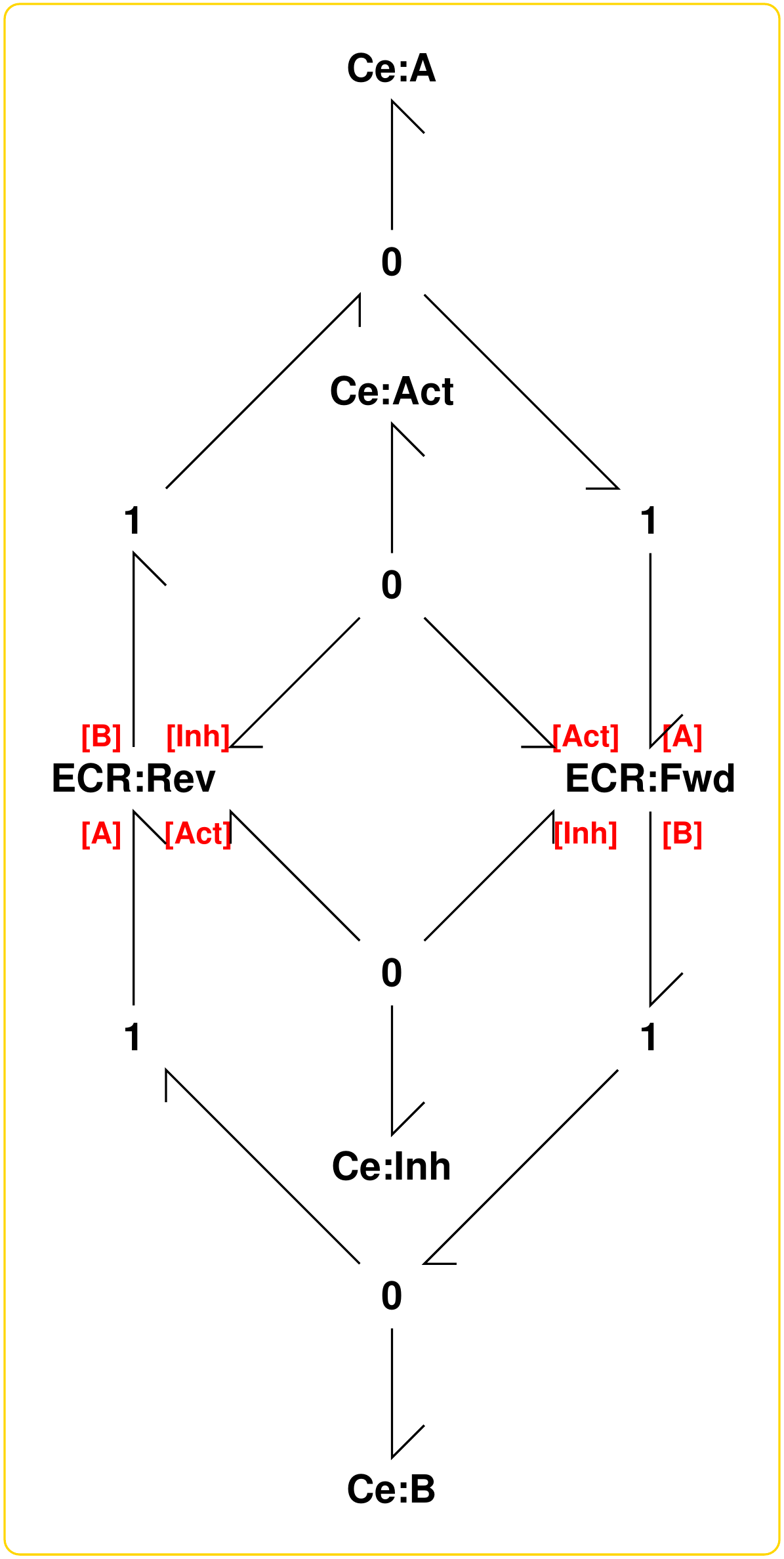}{Cyclic flow modulation (CFM)}{0.3}
  \SubFig{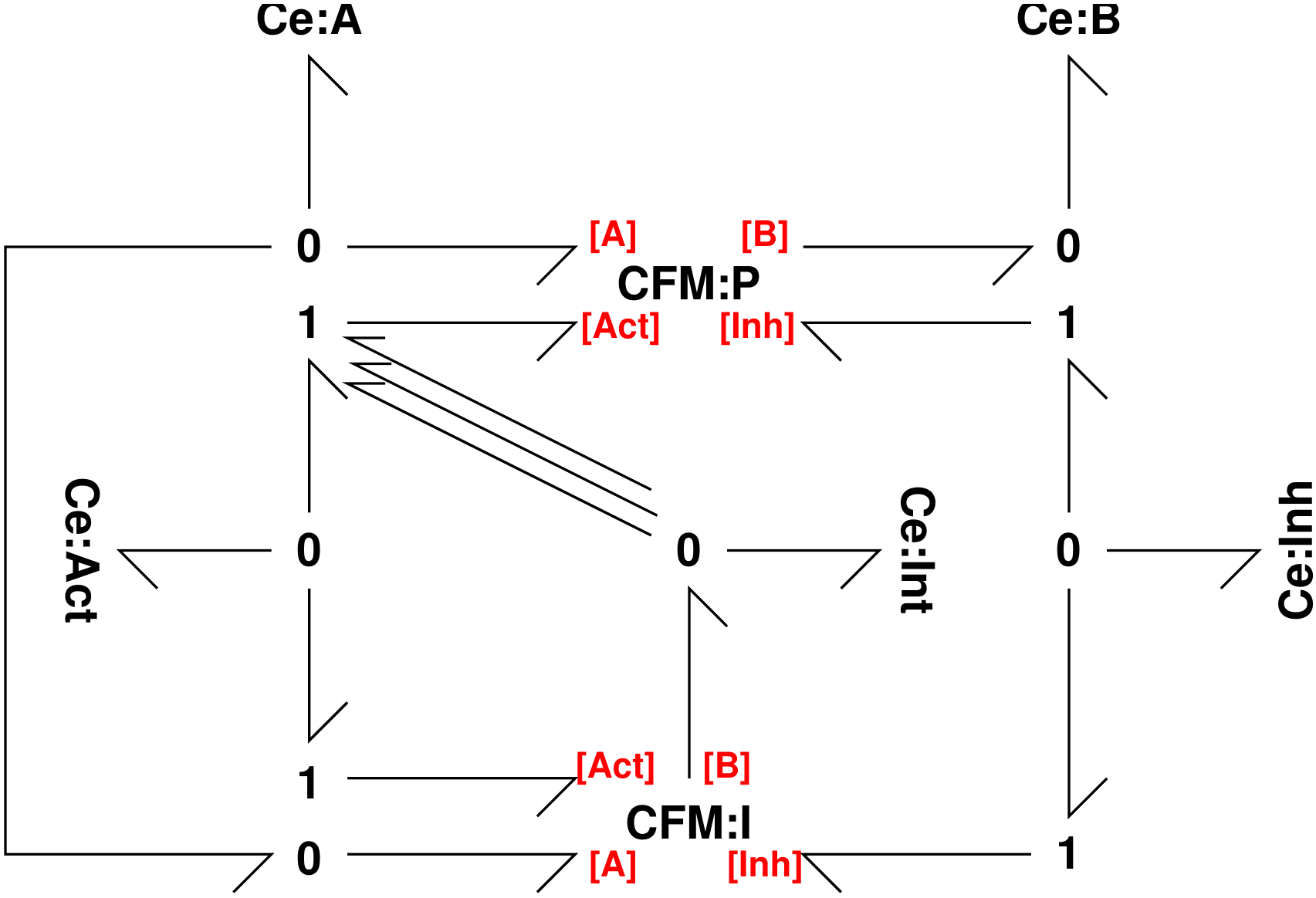}{Integral action (CFMI)}{0.5}
  \caption{Cyclic flow modulation.  (a) The two components
    \BGL{ECR}{Fwd} and \BGL{ECR}{Rev} are instances of the modulated
    ECR of Figure \ref{subfig:ECR_abg}. The \BCe{A} component
    represents both the substrate of \BGL{ECR}{Fwd} and the product of
    \BGL{ECR}{Rev}; the \BCe{B} component represents both the
    substrate of \BGL{ECR}{Rev} and the product of
    \BGL{ECR}{Fwd}. Component \BCe{Act} both activates \BGL{ECR}{Fwd}
    and inhibits \BGL{ECR}{Rev}; component \BCe{Inh} both inhibits
    \BGL{ECR}{Fwd} and activates \BGL{ECR}{Rev}.
    (b)\BGL{CFM}{P} 
    gives proportional (P) action
    whereas \BGL{CFM}{I} gives integral (I) action by driving the
    species \BCe{Int} which activates \BGL{CFM}{P}. To exemplify
    strong activation, three activation bonds are used.
  }
  \label{fig:CFM}
\end{figure}

``\emph{The parallel existence of two irreversible reactions is of the
greatest importance in metabolic regulation: it means that the
direction of flux between two metabolites is determined by
differential regulation of the activities of the two enzymes}''
\cite{Cor12}.
%
A bond graph  interpretation of this mechanism appears
in Figure \ref{subfig:CFM_abg} and this will be used as the basis replacing
the \textbf{CON} component in the bond graph  feedback loop of Figure
\ref{subfig:fb_abg} by a more sophisticated control actuator.

The use of such cyclic flow modulators is motivated by the pair of key metabolic
reactions discussed by \citet{Cor12}:
\begin{align}
  \ch{F6P + ATP &<>[PFK] F16P + ADP}\label{eq:F16P}\\
  \ch{F16P + H2O &<>[PBP] F6P + Pi}
\end{align}
This pair of reactions can be related to the CFM bond graph of Figure
\ref{subfig:CFM_abg} (with reference to Figure \ref{subfig:ECR_abg}) as
follows.
The enzyme corresponding to \BGL{ECR}{Fwd} is \ch{PFK}
(phosphofructokinase) and the enzyme corresponding to \BGL{ECR}{Rev}
is \ch{FBP} (fructose biphosphatase). As noted in
Fig. \ref{fig:ECR_abg}, the generic substrate \ch{A} corresponding to
\BCe{A} is replaced by \ch{F6P} (fructose-6-phosphate) and the generic
product \ch{B} corresponding to \BCe{B} is \ch{F16P}
(fructose-1,6-biphosphate). Within the module \BGL{ECR}{Fwd} (see Fig.
\ref{subfig:ecr_abg}), \BCe{F} corresponds to \ch{ATP} (Adenosine
triphosphate) and \BCe{G} corresponds to \ch{ADP} (Adenosine
diphosphate); within the module \BGL{ECR}{Rev} (see Fig.
\ref{subfig:ecr_abg}), \BCe{F} corresponds to \ch{H2O} and \BCe{G}
corresponds to \ch{Pi} (inorganic phosphate).
This procedure is based on the fact that \Ce components within a
module, that are not exposed as ports, represent a different species for
each instance of the module.

Species which activate PFK and inhibit FBP include \ch{AMP} (Adenosine
monophosphate) and \ch{F26P} (fructose-2,6-phosphate); species which
inhibit PFK and activate FBP include \ch{ATP} and \ch{Cit} (citrate).

\begin{figure}[htbp]
  \centering
  \SubFig{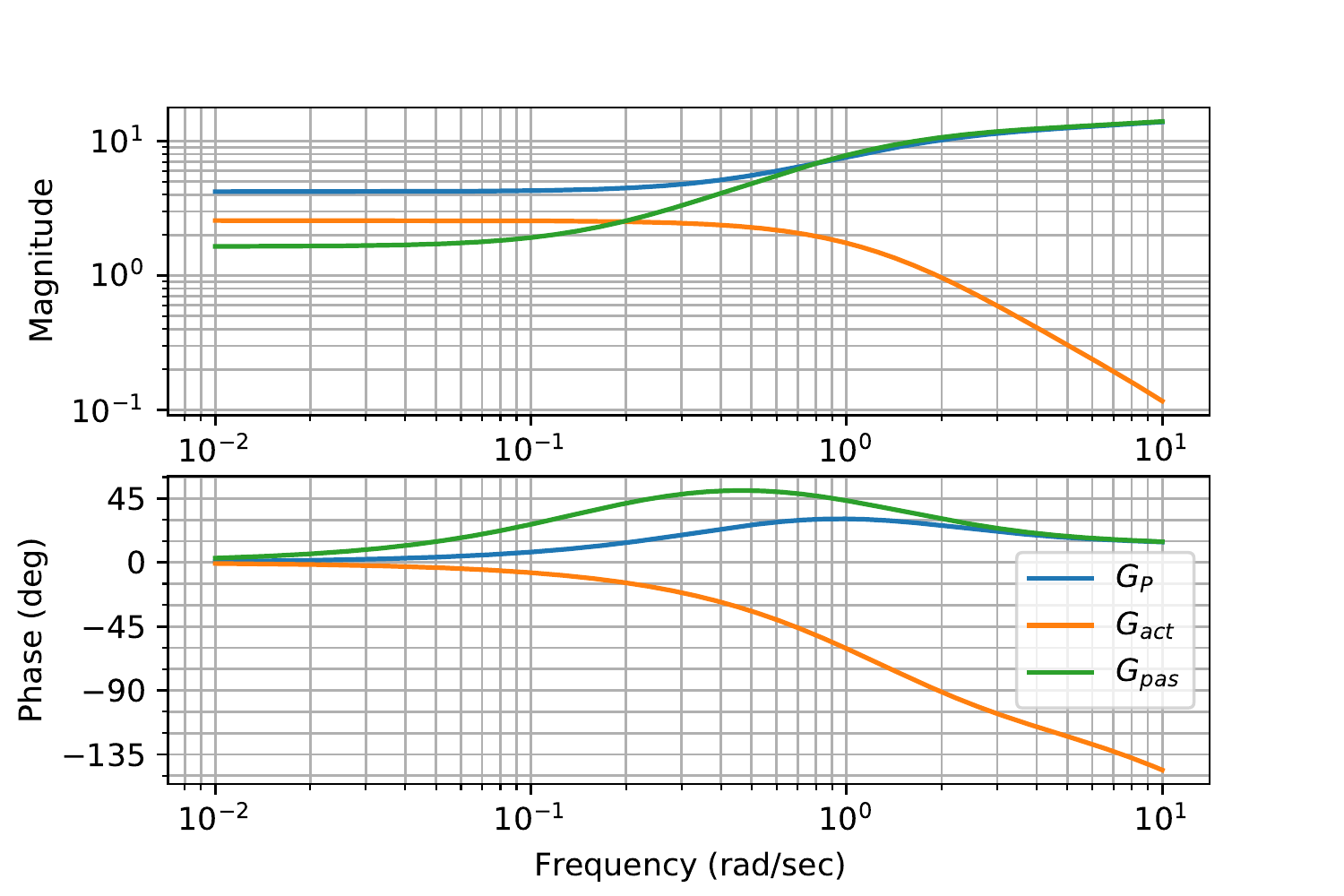}{Components of $G_P(s)$}{0.9}
  \SubFig{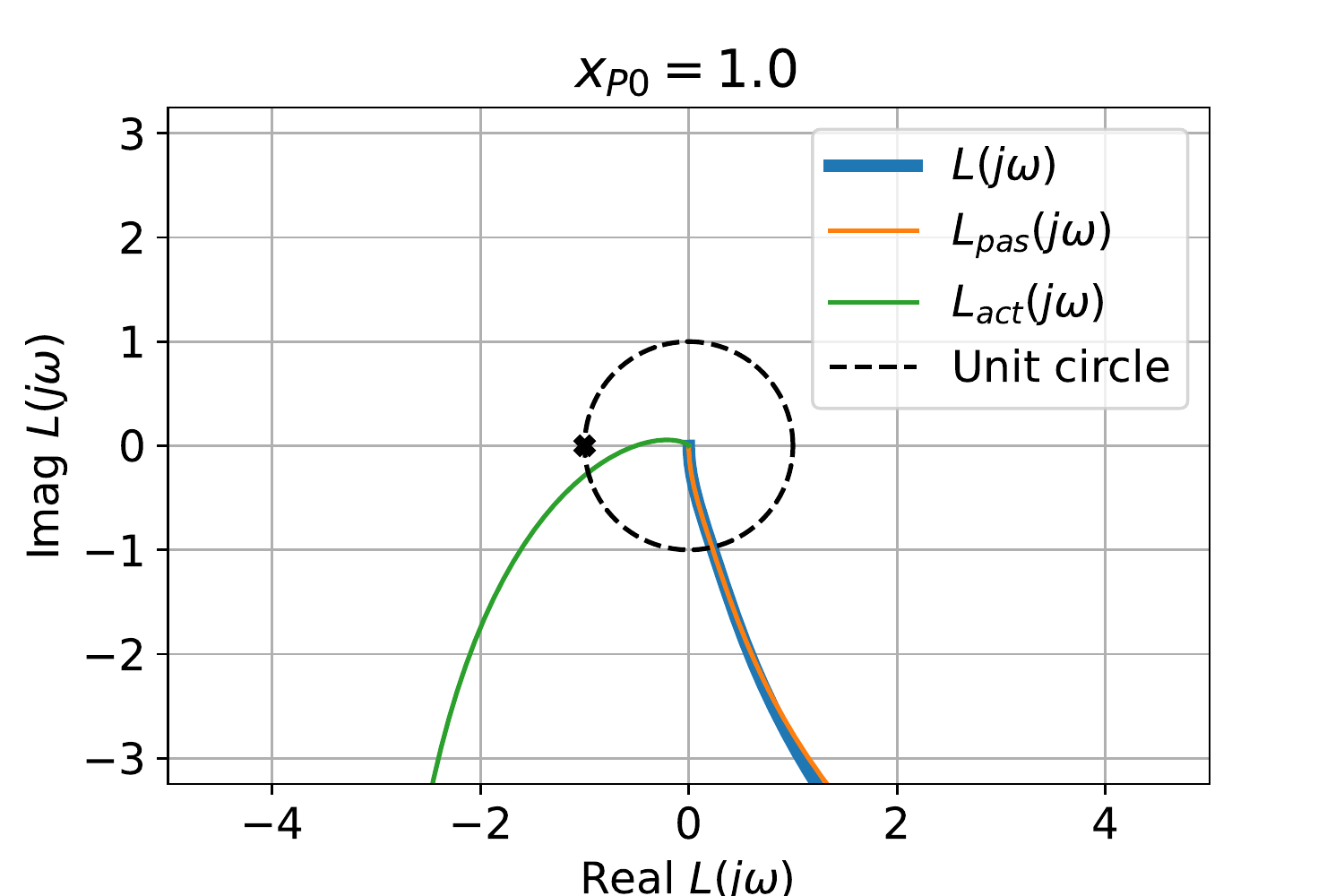}{Open-loop frequency response}{0.45}
  \SubFig{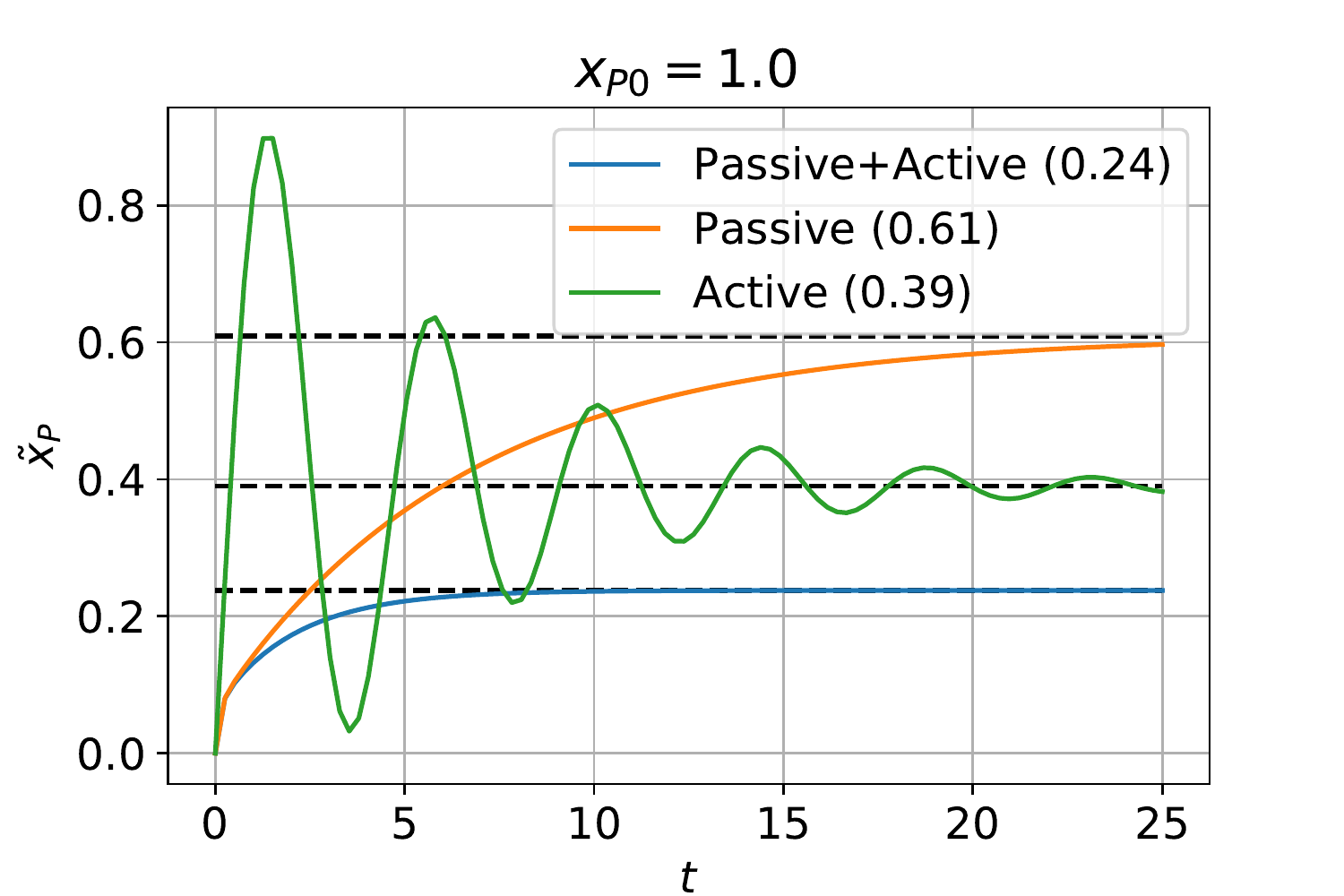}{Closed-loop step response}{0.45}
  \caption{CFM: Split-loop analysis.
    The controller \textbf{CON} and
    system \textbf{SYS} of Fig. \ref{subfig:fb_abg} are replaced by
    the CFM module \textbf{CFM} of Fig. \ref{subfig:CFM_abg} and the pathway module of
    Fig. \ref{subfig:Path_abg} respectively.
    Detailed comments and parameters are given in Figure
    \ref{fig:ECR_analysis}; the low frequency gain is higher leading
    to a lower steady-state error. Once again, the passive term
    $G_{pas}$ stabilises the high-gain active control term $G_{act}$.
  }
  \label{fig:P_analysis}
\end{figure}
This section examines the effect of  replacing the ECR based control of the
feedback loop of Figure \ref{subfig:fb_abg} by a CFM based
controller. The ECR module of  Figure \ref{fig:ECR_abg} has four
visible chemostats \BCe{A}, \BCe{B}, \BCe{Act}, and \BCe{Inh} the
latter three of which are used as ports ([B], [Act], [Inh]) in the
feedback loop of Figure \ref{subfig:fb_abg}; the same chemostats
(\BCe{A}, \BCe{B}, \BCe{Act}, and \BCe{Inh})
are
visible in the CFM module of Figure \ref{subfig:CFM_abg} and can be used as
ports in the same way.
Thus the CFM module can directly replace the
ECR module in the feedback loop of Figure \ref{subfig:fb_abg}
(which was analysed in Figure \ref{fig:ECR_analysis}); this
CFM-based feedback loop is analysed in Figure \ref{fig:P_analysis}.

The linearised response of the ECR and CFM are similar for the
parameters chosen. However, there is a significant difference: the CFM
controller is bidirectional, the ECR is not.
In both cases, the constant low-frequency gain corresponds to the
\emph{proportional} (P) controller of classical control.
In contrast, the next section shows that two CFMs can be combined to
give the classical proportional+integral by endowing the controller
with \emph{integral action}.

\subsection{Integral action}
\label{sec:feedback-control}
\emph{Integral action} is an important concept in classical control
theory \cite{AstMur08} and endows a control system with zero
steady-state error.
In section \emph{3.5 Integral feedback in energy metabolism: the
forgotten side reaction} of their paper \citet{CloWel10} discuss the
role of  \ch{F26P} (fructose-2,6-biphosphate), a strong activator of \ch{PFK}
(phosphofructokinase). In particular,
\ch{F26P}
interconverts with \ch{F6P} (fructose-6-phosphate) via the reaction
cycle:
\begin{align}
  \ch{F6P + ATP &<>[PFK2] F26P + ADP}\label{eq:F26P}\\
  \ch{F26P + H2O &<>[F26BP] F6P + Pi}
\end{align}
catalysed by the enzymes \ch{PFK}2 (phosphofructokinase-2) and
\ch{F26BP} (fructose-2,6-biphosphatase).  The species which
simultaneously activate PFK2 and inhibit F26BP include \ch{AMP} and
\ch{F6P}. Hence this pair of reactions is a further example of Cyclic
Flow Modulation (CFM).

Moreover, the PFK CFM and the PFK2 CFM strongly interact:
  the PFK CFM is positively modulated by the product of the PFK2 CFM:
  \ch{F26P} and
  both are positively modulated by \ch{AMP}.

Figure \ref{subfig:PI_abg} gives the bond graph abstraction of the two
interacting cycles. \BGL{CFM}{P} corresponds to the \ch{PFK}-based CFM
giving proportional (P) action
whereas \BGL{CFM}{I} corresponds to the \ch{PFK}2-based CFM and, as
will be seen, gives integral (I) action. Within each
CFM, the interpretation of the species is the same except that the
product \ch{B} of  \BGL{CFM}{I} corresponds to \ch{F26P} rather than
\ch{F16P}. The component \BCe{Int} corresponds to the product
\ch{F26P} which then activates \BGL{CFM}{P}. For illustration, and to
emphasise the strong activation effect, three bonds represent the
activation effect.

\begin{figure}[htbp]
  \centering
  \SubFig{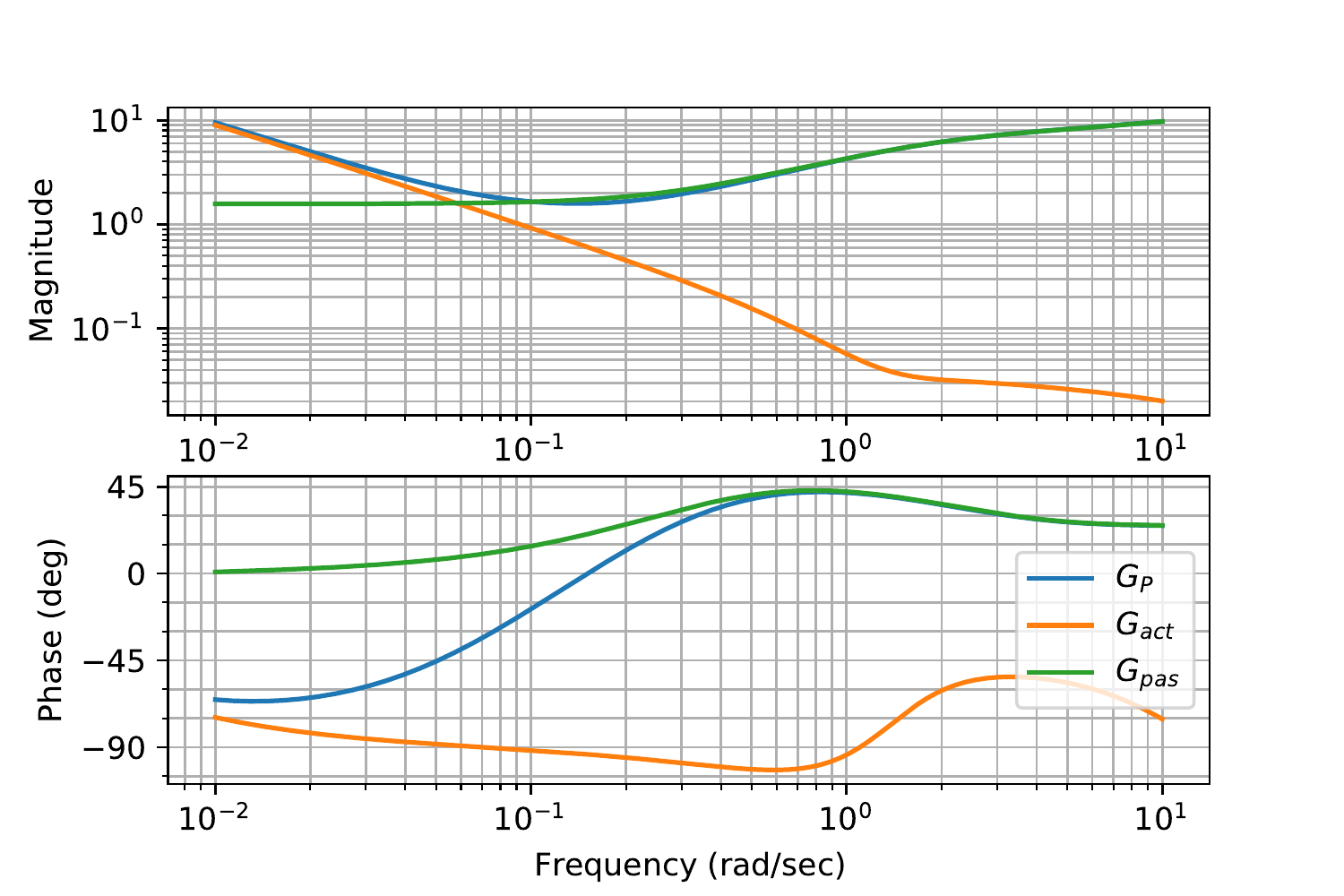}{Components of $G_P(s)$}{0.9}
  \SubFig{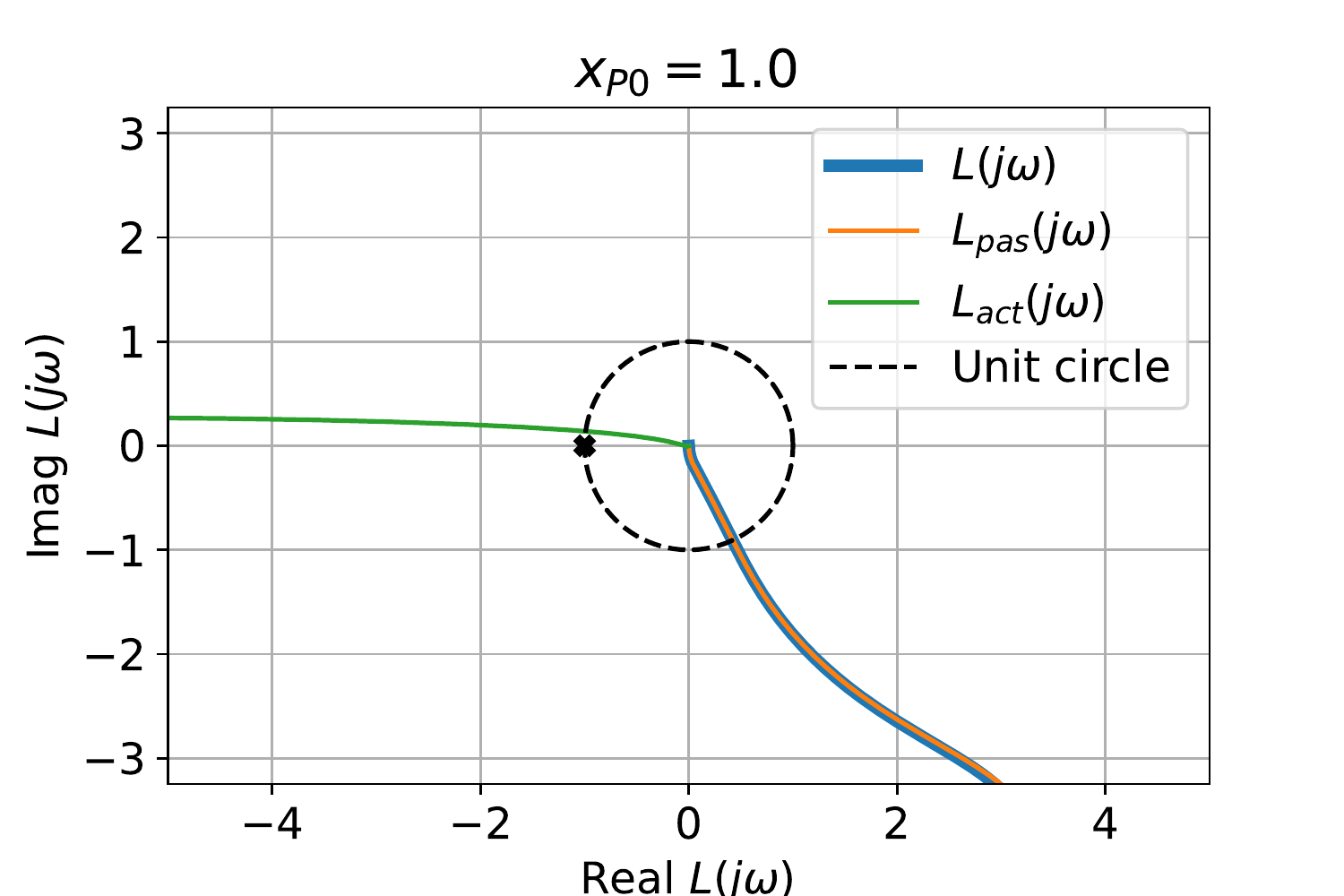}{Open-loop frequency response}{0.45}
  \SubFig{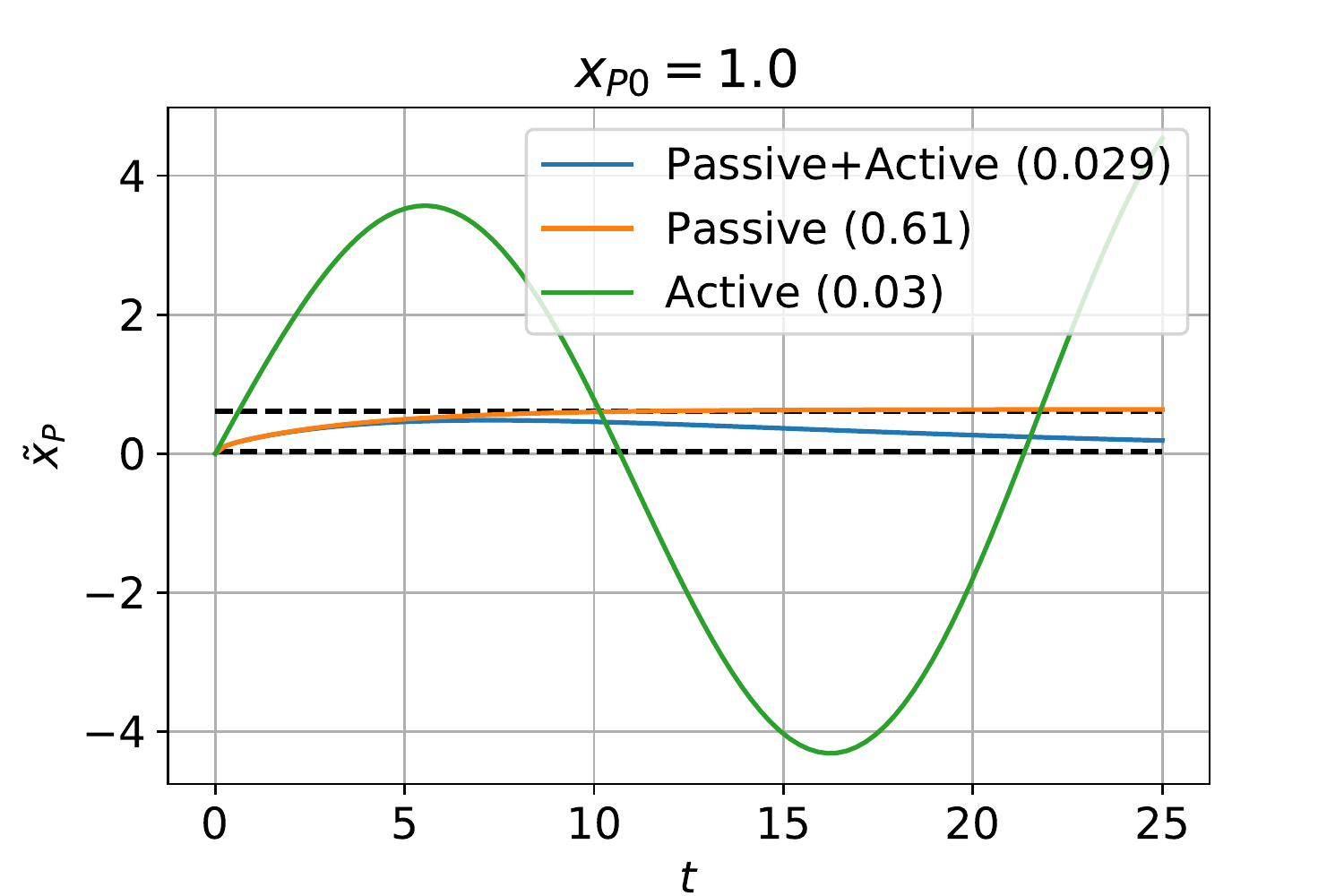}{Closed-loop step response}{0.45}
  \caption{CFMI: Split-loop analysis.
    The controller \textbf{CON} and
    system \textbf{SYS} of Fig. \ref{subfig:fb_abg} are replaced by
    the CFM module \textbf{CFMI} of Fig. \ref{subfig:PI_abg} and the pathway module of
    Fig. \ref{subfig:Path_abg} respectively.
    Detailed comments and parameters
    are given in
    Figure \ref{fig:ECR_analysis}. Compared to the CFM controller
    response of Figure \ref{fig:P_analysis}, the low-frequency gain of
    the active term $G_{act}$ rises as frequency decreases; this is
    the behaviour expected of an integrator. However, as the
    integrator is not perfect, the gain is not infinite at
    $\omega=0$. This approximate integral effect gives a lower
    steady-state error than the CFM controller whilst the
    passive term $G_{pas}$ continues to act to give a damped response.
    As the phase of $L_{act}(\jw_c)$ is below $-180^\circ$ at the
    critical frequency $\omega_c$ at which magnitude
    $|L_{act}(\jw_c)|=1$, the closed-loop system corresponding to the
    active part of the controller is unstable.  }
  \label{fig:PI_analysis}
\end{figure}


\begin{figure}[htbp]
  \centering
  \SubFig{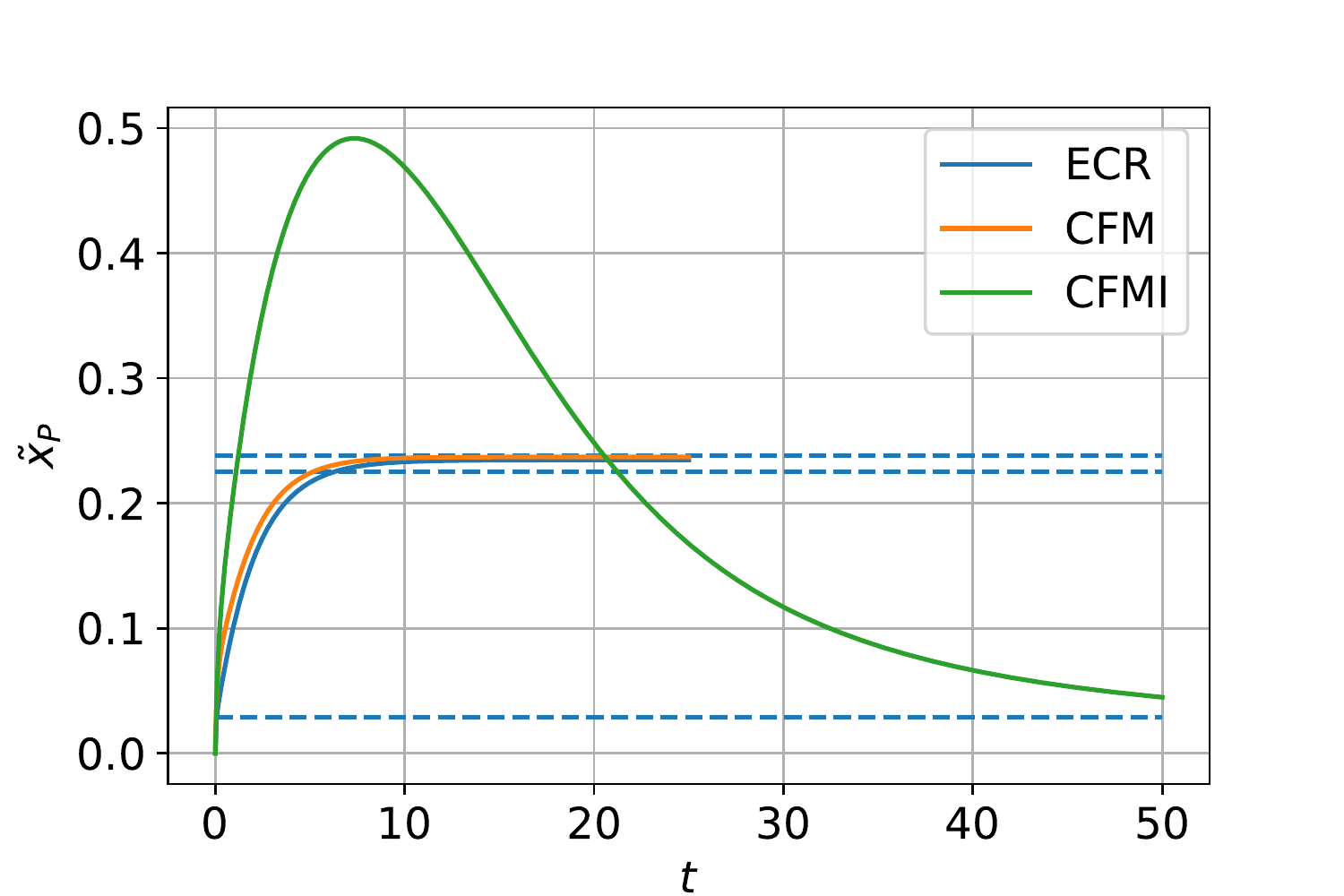}{Nonlinear}{0.75}
  \SubFig{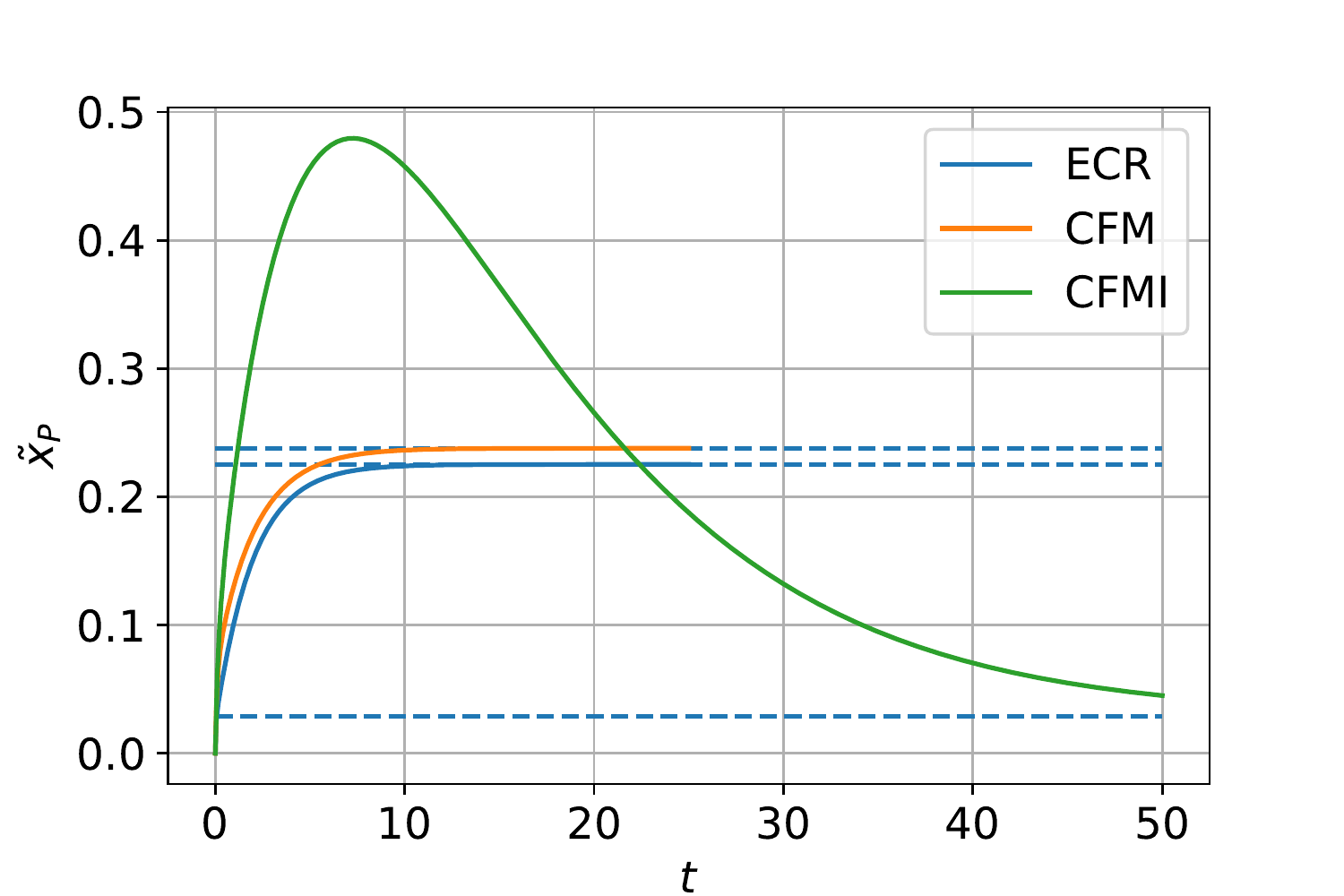}{Linearised}{0.75}
  \caption{Controller comparison.
    (a) The response of the product concentration deviation from
    steady-state $\xt_P = \xx_P - \xb_P$ to a unit step deviation in
    disturbance concentration $\xt_D$ for each of the 
    three controllers ECR, CFM and CFMI is plotted against normalised time $t$ together with the steady state
    disturbance gains $g_D$ (\ref{eq:g_D}) as dashed lines. Thus the CFMI
    controller is the best of the three in reducing the steady-state
    error.
    (b) The linearised responses are shown for comparison; they are
    close to the nonlinear responses and asymptotically match the
    disturbance gains $g_D$ (\ref{eq:g_D}).
  }
  \label{fig:compare}
  \end{figure}

This section examines replacing the CFM based control within the
feedback loop of Figure \ref{subfig:fb_abg} by a CFMI based
controller. As in the case of CFM, the same chemostats as ECR are
visible in the CFMI module of Figure \ref{subfig:PI_abg} and can be used as
ports in the same way. Thus the CFMI module can directly replace the
ECR module in the feedback loop of Figure \ref{subfig:fb_abg}
analysed in Figure \ref{fig:ECR_analysis}; this CFMI-based feedback
loop is analysed in Figure \ref{fig:PI_analysis}.
Compared to the CFM controller response of Figure
\ref{fig:P_analysis}, the low-frequency gain of the active term
$G_{act}$ rises as frequency decreases; this is the behaviour expected
of an integrator. However, as the integrator is not perfect, the gain
is not infinite at $\omega=0$; but this approximate integral effect
gives a significantly lower steady-state error than the CFM controller
whilst the passive term $G_{pas}$ continues to act to give a damped
response.
The disturbance response of the three controllers (using nonlinear
simulation) is shown in Figure~\ref{subfig:PI_nstep}; the CFMI
controller has a substantially smaller steady-state disturbance error
than the other two.
For comparison, the corresponding linearised responses appear in
Figure~\ref{subfig:PI_step}; they are close to the nonlinear response
and, as expected, are asymptotic to the the steady state disturbance
gains $g_D$ (\ref{eq:g_D}).

Performance depends on both controller structure and controller
parameters.
Figure~\ref{fig:compare} illustrates that the CFMI structure is superior
to both the ECR and CFM structures.
As discussed in Figure~\ref{fig:ECR_abg}, the generic species \ch{F}
and \ch{G} provide the energy driving the controller reactions;
for example, in reactions (\ref{eq:F16P}) and (\ref{eq:F26P}), \ch{F}
represents \ch{ATP} and \ch{G} represents \ch{ADP}.
Thus the concentrations $x_F$ and $x_G$ together with the parameters
$K_F$ and $K_G$ (equation (\ref{eq:CR_C})) are crucial to control
performance.  In particular, the effective controller gain increases
with $K_F/K_G$ and the simulated values correspond to the values of
$K_F$ and $K_G$ given in Figure~\ref{fig:step}; thus large $K_F/K_G$
improves controller performance at the expense of increased energy
consumption.
An analysis of this performance/energy tradeoff is beyond the scope of
this paper.

The similarity of linear and nonlinear responses validates the use of
linear design approaches in this nonlinear context. It would be
interesting to use the frequency response methods of this paper to
investigate issues such as robustness to the changes in parameters of
the linearised system due to both changes in the equilibria and the
parameters of the underlying nonlinear system.



\subsection{Steady-state values}
\label{sec:steady-state-values}
\begin{figure}[htbp]
  \centering
  \SubFig{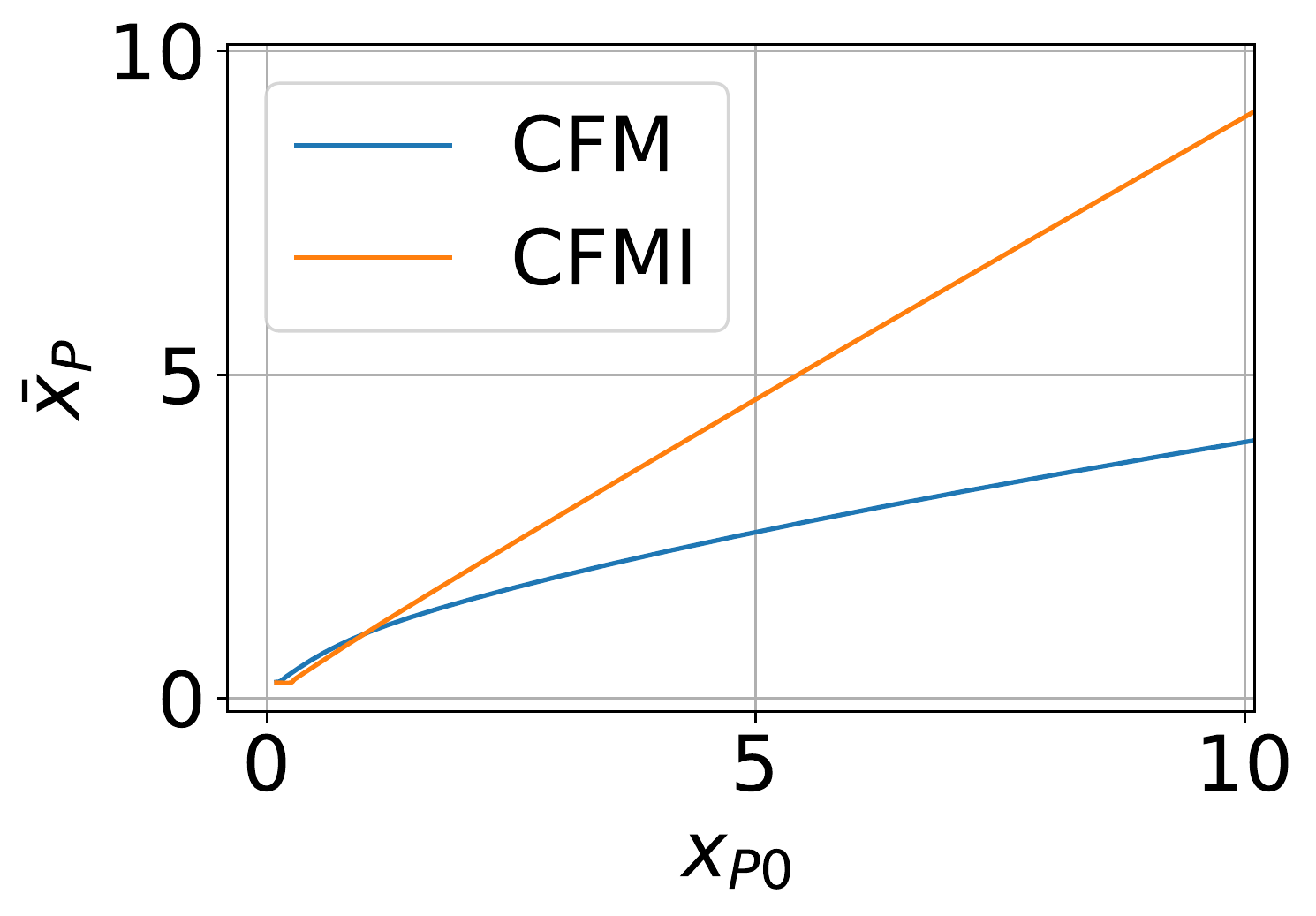}{$\xb_P$}{0.4}
  \SubFig{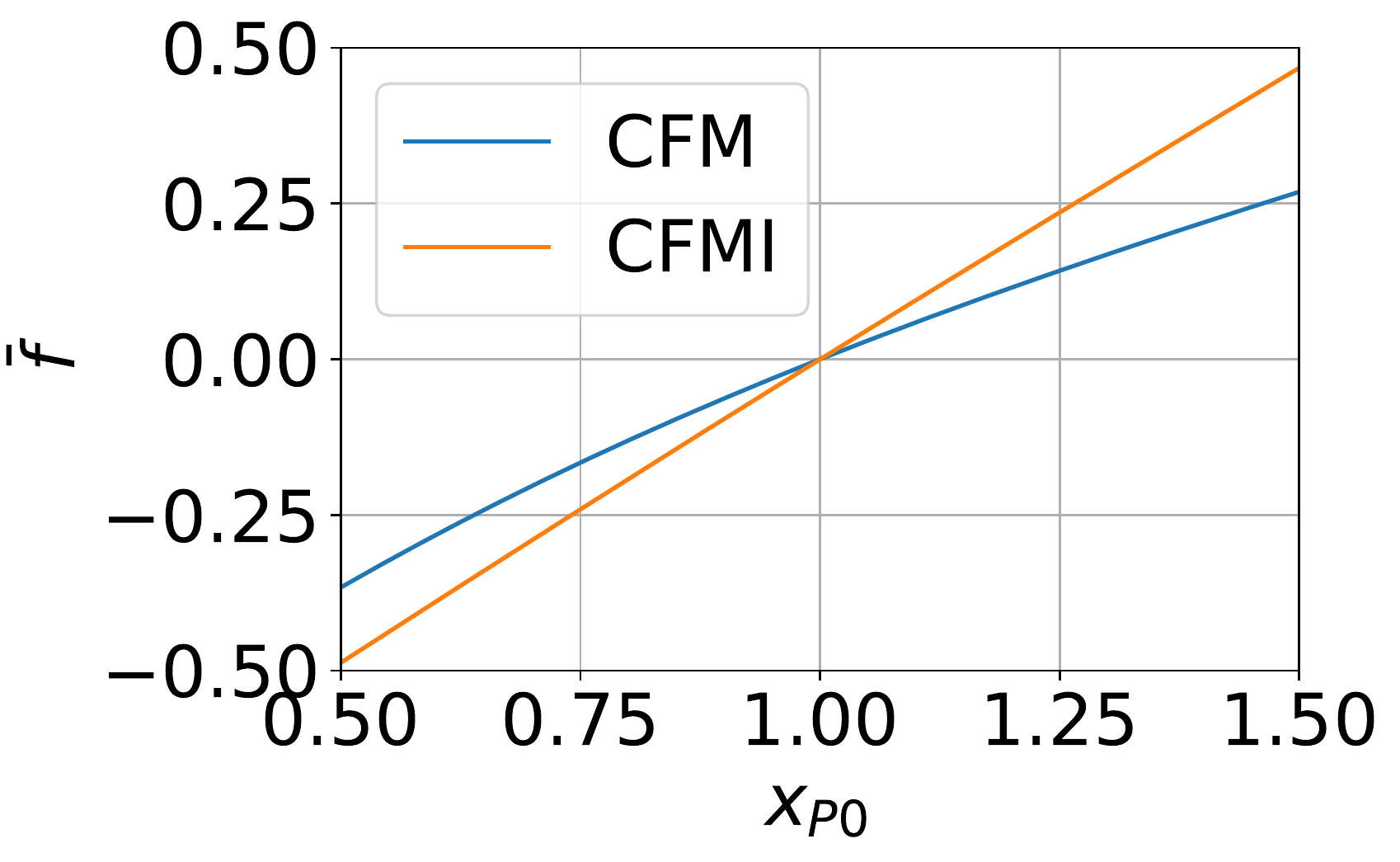}{$\fb_P$}{0.45}
  \caption{Steady-state.
    (a) The steady-state product state $\xb_{P} \approx x_{P0}$ for
    the CFM and CFMI
    controllers.
    (b) The steady-state product flow $\fb_P$ is bidirectional in the
    case of the two CFM-based controllers; this is not possible for
    the ECR controller.
  }
  \label{fig:steady-state}
\end{figure}
In the examples so far, the activation chemostat of Figure
\ref{subfig:fb_abg} is defined by a unit state $x_{P0}=1$. By
analogy with the classical feedback loop of Figure
\ref{subfig:feedback}, it would be expected that $x_{P0}$ would play a
similar role to $w$. Figure \ref{subfig:x_ss} indicates that this is
approximately true for the CFMI control: $\xb_{P}\approx x_{P0}$.
Furthermore, varying $x_{P0}$ changes the steady-state product
flow. In this case, as the disturbance reaction gain is
$\kappa_{rd}=1$ the product flow $\fb_P = x_P-x_d = x_P-1$. One of the
benefits noted for CFM control at the beginning of \S~\ref{sec:CFM} is
that bidirectional product flow is possible: Figure \ref{subfig:f_ss}
illustrates this for the CFM and CFMI controller; it is not possible
for the ECR controller.


\section{Conclusion}\label{sec:conclusion}
Network thermodynamic modelling via bond graphs has been amalgamated
with classical control theory.
This means that physically-based models of biomolecular systems, which
automatically include features -- such as retroactivity -- which are
not ideal from a control systems viewpoint, can be directly analysed
using the transfer function based methods of control theory and the
resultant designs are thus biologically feasible.
In particular, the dual roles of active and passive feedback have been
analysed: active feedback gives good steady state performance whereas
passive feedback provides stabilisation.
  
In this context, cyclic flow modulation (CFM) has been motivated by the
phosphofructokinase-fructose biphosphatase cycle of the metabolic pathway
and shown to have a modular bond graph representation.
The methods of this paper could be used in two ways: to \emph{analyse} the CFM-based control
systems arising from evolution and to \emph{synthesise} new cellular
control systems. For example,
CFM could be used to build biological analogues of the proportional (P)
and proportional+integral (PI) controllers of classical control
theory, as well as allowing bidirectional flow modulation. Such
non-ideal controllers can then be analysed by the methods of this
paper and, as models of biological systems, can be directly
implemented as biological systems.
  
Future work will include building an \emph{energy-based} model of
metabolism with \ch{AMP} feedback and mitochondrial transduction using
existing energy-based models \citep{Gaw17a,GawCudCra20}.
  
An important potential result of combining control theory with
energy-based modelling is to identify performance/energy trade-offs
and this is the subject of current research. This tradeoff is
important to both evolutionary theory~\cite{Lan20} and synthetic
biology~\cite{DelCheWad20}.

  \section{Acknowledgements}
  Peter Gawthrop would like to thank the Melbourne School of
  Engineering for its support via a Professorial Fellowship, and
  Edmund Crampin and Michael Pan for help, advice and encouragement.
  The suggestions of anonymous reviewers were instrumental in
  improving this paper.


\end{document}